\documentclass[pdflatex,sn-nature]{sn-jnl}

\usepackage{graphicx}%
\usepackage{multirow}%
\usepackage{amsmath,amssymb,amsfonts}%
\usepackage{amsthm}%
\usepackage{mathrsfs}%
\usepackage[title]{appendix}%
\usepackage{xcolor}%
\usepackage{textcomp}%
\usepackage{manyfoot}%
\usepackage{booktabs}%
\usepackage{algorithm}%
\usepackage{algorithmicx}%
\usepackage{algpseudocode}%
\usepackage{listings}%
\usepackage{natbib}
\usepackage{xurl}

\theoremstyle{thmstyleone}%
\theoremstyle{thmstyletwo}%

\theoremstyle{thmstylethree}%

\begin{document}

\title[Article Title]{The Neuromorphic Supremacy}


\author[1,2]{\fnm{Yuliya} \sur{Tsybina}}\email{lotarevaj@gmail.com}

\author*[3,4]{\fnm{Ivan Y.} \sur{Tyukin}}\email{ivan.tyukin@kcl.ac.uk}

\author[3]{\fnm{Alexander N.} \sur{Gorban}}\email{alexander.gorban@kcl.ac.uk}

\author[1,2]{\fnm{Victor} \sur{Kazantsev}}\email{kazantsev@neuro.nnov.ru}

\author*[4]{\fnm{Dianhui } \sur{Wang}}\email{dh.wang@deepscn.com}

\author*[1,2]{\fnm{Susanna} \sur{Gordleeva}}\email{gordleeva@neuro.nnov.ru}

\affil*[1]{\orgdiv{Department of Neurotechnology}, \orgname{Lobachevsky State University of Nizhny Novgorod}, \orgaddress{\street{Gagarin Ave.}, \city{Nizhny Novgorod}, \postcode{603022}, \country{Russia}}}

\affil[2]{\orgdiv{Neuromorphic computing center}, \orgname{Neimark University}, \orgaddress{\street{Nartova}, \city{Nizhny Novgorod}, \postcode{603081}, \country{Russia}}}

\affil*[3]{\orgdiv{Department of Mathematics}, \orgname{King's College London}, \orgaddress{\street{Strand}, \city{London}, \postcode{ WC2R 2LS}, \country{United Kingdom}}}

\affil*[4]{\orgname{ International Joint Laboratory of AI for Industry, Qingdao University of Science and Technology}, \orgaddress{\street{99 Songling Rd}, \city{Qingdao}, \postcode{266101},  \country{China}}}

\abstract{Live neural systems demonstrate remarkable capabilities to learn new behavior and patterns from mere few examples and are known to operate robustly under severe sensory noise. These capabilities, however, remain largely out of reach for modern artificial neural networks, including deep learning models. We show that this gap can be bridged by embedding novel genuine neuromorphic circuits into conventional artificial neural network architectures. These circuits comprise astrocytic modulation and spiking dynamics inherent to biological neural structures. Tested across standard benchmarks representing tasks of varying complexity, the hybrid models achieve high accuracy from  few training examples per class and sustain high performance under occlusion and impulse noise that cause performance collapse in standard models without neuromorphic adaptation. We term this phenomenon neuromorphic supremacy — a regime in which architectures grounded in neurobiology decisively outperform classical deep learning, pointing toward a principled foundation for perception in embodied AI systems operating in noisy, data-scarce environments.}


\maketitle

\section{Introduction}\label{sec1}

Since the early days of Artificial Intelligence (AI), understanding how live neural systems function and evolve over time has been a key and major pillar behind its success, supporting the development and creation of AI architectures capable of learning from empirical data and experience. Starting from simple Rosenblatt's perceptron \cite{rosenblatt1958perceptron}, these AI systems developed into a thriving eco-system of nature-inspired computational models with different architectures including classical feed-forward neural networks, convolutional neural networks \cite{lecun1998convolutional}, transformers \cite{vaswani2017attention}, and their various combinations. 

Modern AI systems adhering to these basic architectures  have already shown exceptional capabilities. Examples of their success include human-level performance in the analysis of medical images \cite{hernstrom2025masai}, excellent skills in solving mathematical \cite{deepmind2025imo} and science questions \cite{abramson2024alphafold3}, \cite{merchant2023gnome} and a recognition by the 2024 Nobel Prize committee for predicting complex protein structures \cite{nobel2024chemistry}. Yet, these highly capable AI models are suffering from fundamental limitations: 

\begin{itemize}
\item[] {\it They can be highly sensitive to minute and imperceivable changes of input data \cite{bastounis2026survey}, \cite{bastounis2023boundaries} and often require large datasets for training \cite{hoffmann2022chinchilla}, \cite{udandarao2024exponential}. In absence of sizable datasets their efficiency degrades.}  
\end{itemize}

Unlike modern AI models, biological neural systems  exhibit extraordinary robustness to noise and do not require hundreds or thousands of examples to learn new classes, at least in vision tasks \cite{wood2013newborn}, \cite{wood2020oneshot}, \cite{geirhos2018generalisation}, \cite{geirhos2020shortcut}. The question is: why do we observe these discrepancies, given that modern AI models are designed to mimic and replicate computations occurring in live neural systems? 

Answering this question requires an understanding of key determinants powering continuous progress in AI. These are hardware improvements, new algorithms, and new AI architectures. Indeed, over the span of nearly half-a-century AI systems benefited from a continuous and drastic growth in compute power \cite{hennessy2019goldenage}. Yet, for many computationally hard problems, gains from algorithmic improvement have exceeded gains from hardware improvement over recent decades \cite{hernandez2020efficiency}, \cite{nrc2010future}, \cite{sherry2021algorithmic}. However, acknowledging the importance of hardware and algorithms, the most consequential advances in deep learning over the last few decades have been predominantly architectural rather than algorithmic. Indeed, while the foundational principle of gradient-based learning via backpropagation has remained constant, transformative gains have come from architectural innovations: from fully connected networks to convolutional architectures that exploit spatial locality \cite{lecun1998convolutional}, and subsequently to transformers that capture long-range dependencies through self-attention mechanisms \cite{vaswani2017attention}. Notably, each such successive architectural change brings AI models closer to the organizational principles observed in biological neural systems. MLPs abstract the hierarchical, layered connectivity observed in cortex \cite{felleman1991distributed}.  Convolutional networks mirror the hierarchical receptive field structure of the visual cortex \cite{Hubel1962}, while attention mechanisms parallel selective neural processing observed in primate visual systems \cite{Itti2001}.
 
In this paper, we propose the next step in this architectural progression:  explicit incorporation of inherently neuromorphic computational elements \cite{Taherkhani2020, Roy2019} -- astrocytic modulation and spiking dynamics -- into conventional deep learning architectures. Unlike their counterparts in many modern AI models, real neurons elicit dynamic, not static, responses to incoming stimuli. Synaptic weights and weights plasticity in live neurons are dynamic too.  Furthermore, growing evidence implicates astrocytes -- glial cells -- as crucial regulators of neural information processing \cite{Perea2007, Gordleeva2023, Santello2019, MurphyRoyal2023}. These specific neuromorphic mechanisms cannot be easily captured by standard dynamic models such as Long Short-Term Memory (LSTM) \cite{hochreiter1997long}, Hopfield networks \cite{Hopfield1982}, or generic recurrent neural networks.
 
\begin{figure}
\centering
\includegraphics[width=1\textwidth]{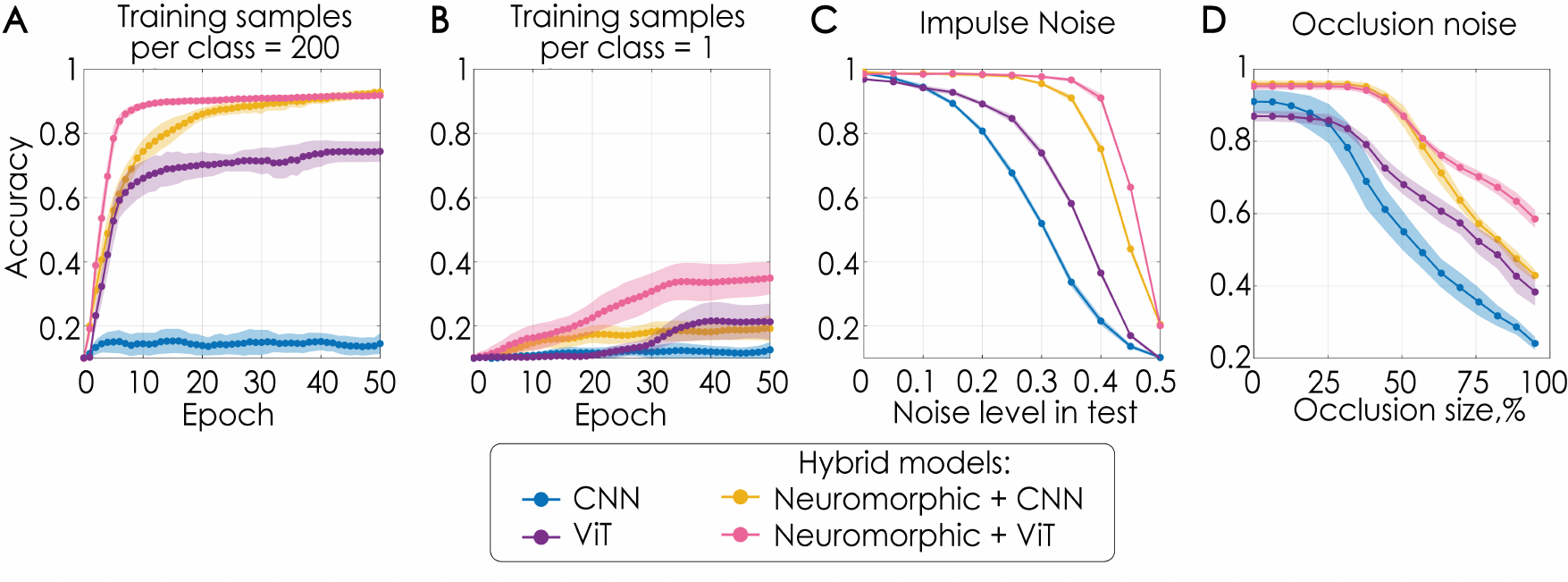}
\caption{Summary of the main results.
(A–B) Comparison of classification accuracy of the four architectures under data scarcity. The models were trained on (A) 200 examples per class (3.3\% of the full MNIST training set) and (B) 1 example per class (0.017\% of MNIST). All test images were corrupted with impulse noise at a level of 0.15.
(C–D) Comparison of classification accuracy of the four architectures: (C) under increasing impulse noise during testing, with models trained at a noise level matched to the test condition; (D) under increasing occlusion size (percentage of target image area).}
\label{Fig1}
\end{figure}

We show that fusing classical state-of-the-art AI architectures with the newly proposed neuromorphic module offers a radical transformation: new hybrid models are significantly more robust to data noise and require substantially less data to learn a task. The results are demonstrated through extensive experiments on classical benchmark datasets of growing complexity: binarized MNIST \cite{lecun1998convolutional} (simplest baseline), Fashion-MNIST \cite{xiao2017fashion} (a substantially harder task), and Omniglot \cite{lake2015human} (a dataset  specifically designed to asses performance of few-shot learning). In absence of the new neuromorphic features and in presence of noise, classical architectures such as convolutional neural networks fail to learn even the simplest task such as handwritten digits from MNIST dataset with several hundreds of training images per class. Vision  transformers behave better, yet even these modern architectures fail when the data becomes scarce and reduces to a mere single example (Fig.~\ref{Fig1}A-B).  At the same time, when equipped with a new neuromorphic module, their performance drastically shots up (Fig.~\ref{Fig1}A-B). In contrast to vanilla state-of-the-art conventional models, our hybrid neuromorphic models show remarkable robustness to various patterns of noise (Fig.~\ref{Fig1}C-D) at both training and test  times. A similar pattern -- our new architecture's substantially reduced sensitivity to noise and to the lack of training data -- persists on fashion-MNIST and Omniglot datasets.  These findings suggest that bringing in more realistic neuromorphic circuits into  computational graphs of classical models has a potential to significantly alter the overall performance of AI systems. The phenomenon discovered and reported in our work resembles the well-known phenomenon of quantum supremacy whereby quantum circuits and algorithms drastically outperform conventional computers in certain tasks \cite{preskill2012supremacy_arxiv}. Here we show that the same may apply to data-driven AI models -- there are certain tasks in which neuromorphic architectures significantly outperform classical AI models \cite{Roy2019}. We call this phenomenon neuromorphic supremacy.  
 
The structure of the paper is as follows. In Section \ref{sec2} we present main findings of the work. Methods are detailed in Section \ref{sec_Met}. Section \ref{sec_discussion} contains a brief discussion of results, and Section \ref{sec_conclusion} concludes the paper.

\section{Results}\label{sec2}

\subsection{A Novel Class of Hybrid Neuromorphic Models}\label{subsec2_1}

\begin{figure}
\centering
\includegraphics[width=1\textwidth]{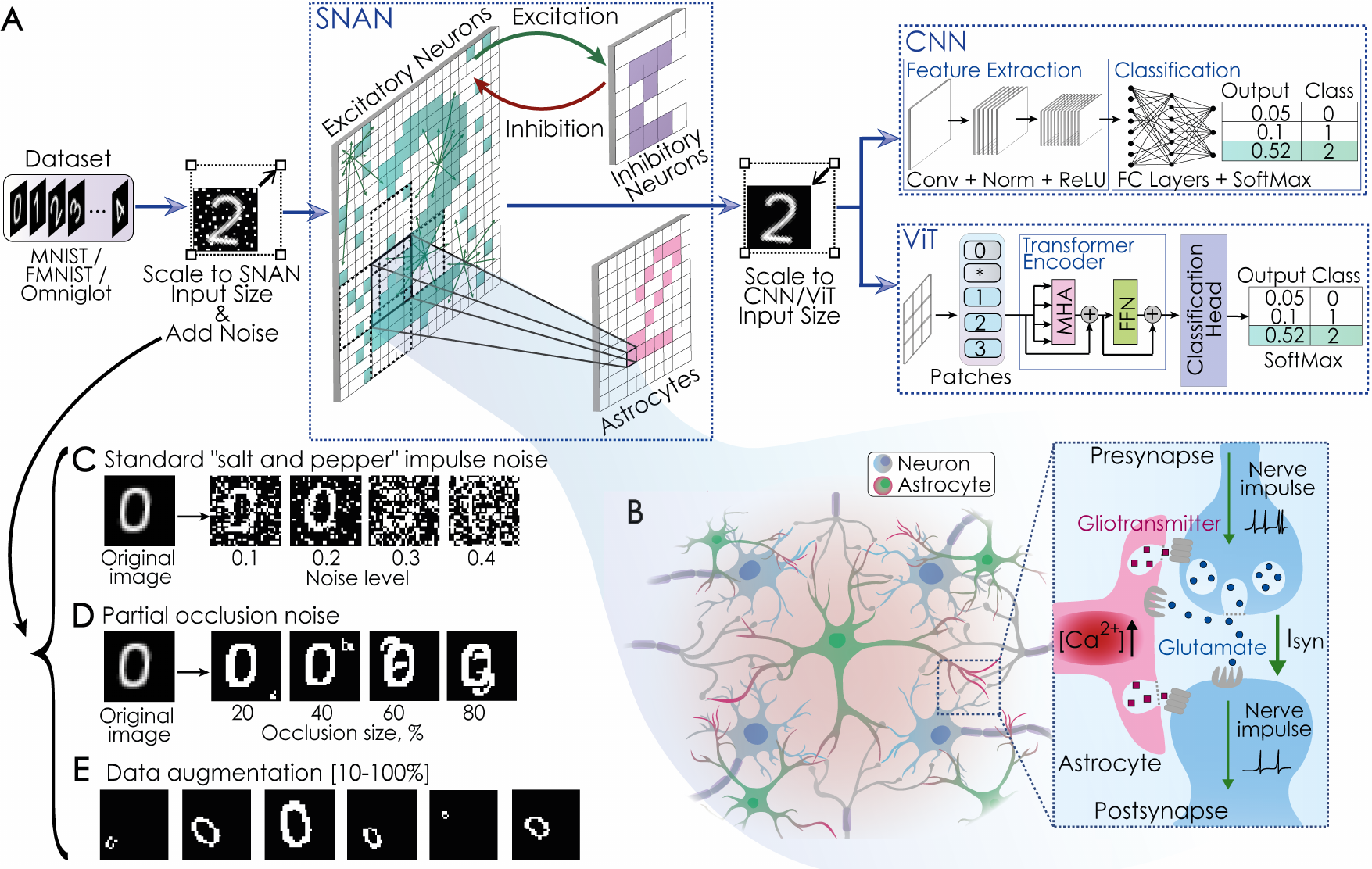}
\caption{(A) Schematic of the Hybrid Neuromorphic Model: Spiking neuron–astrocyte network (SNAN) combined with CNN/ViT. (B) Bidirectional neuron–astrocyte interactions in the SNAN model. (C–E) Examples of images from binarized MNIST: (C) with added standard random ”salt and pepper” impulse noise at varying noise levels; (D) with partial occlusion noise at varying sizes; (E) with random scale-and-shift augmentation applied.}
\label{Fig2}
\end{figure}

We propose a class of hybrid architectures in which a spiking neuron–astrocyte network (SNAN) \cite{Gordleeva2023} serves as a biologically motivated preprocessing stage before conventional classifiers—here, a convolutional neural network (CNN) and a vision transformer (ViT) (Fig.~\ref{Fig2}A). The key feature of the SNAN (see Methods, Section~\ref{sec_Met_SNAN}) is its dual‑timescale computation: fast (millisecond) spiking dynamics encode the stimulus instantaneously, while slow (second‑scale) astrocytic calcium signalling provides contextual modulation, local synchronisation, and one‑shot memory formation Fig.~\ref{Fig2}B). This design fundamentally distinguishes the SNAN from existing recurrent and associative memory architectures. Standard recurrent neural networks (RNNs) \cite{Elman1990} are universal approximators of dynamical systems.  However, they lack specific biological structure; learning the intricate neuron–astrocyte interplay from data alone would be practically infeasible within the class of these models. Long short‑term memory (LSTM) networks \cite{hochreiter1997long} circumvent the vanishing gradient problem via gating mechanisms yet remain single‑timescale recurrent systems built upon the same static neural abstraction. Hopfield networks \cite{Hopfield1982} converge to fixed‑point attractors, implementing content‑addressable memory through energy minimisation, while bidirectional associative memories (BAM) \cite{Kosko1988} extend this principle to hetero‑associative recall—but both remain strictly attractor‑based architectures. In contrast, the SNAN implements a qualitatively different computational principle: fast neural dynamics encode the incoming pattern, whilst slow astrocytic dynamics accumulate  memory traces and perform context‑dependent filtering. The SNAN thus introduces biologically grounded multi‑timescale processing into the computational graph, a capability that existing architectures lack. Because astrocytic modulation relies on learned, stimulus‑specific calcium patterns, the SNAN functions as an adaptive filter that selectively amplifies coherent neural activity and suppresses incoherent noise or occlusion artefacts. This mechanism is fundamentally different from classical denoising pipelines (median, mean, bilateral filters), which operate on fixed local statistics and cannot semantically separate signal from corruption (examples of noise and occlusion are shown in Fig.~\ref{Fig2}C,D).

We systematically compare four SNAN integration strategies, which differ in whether the SNAN is applied only at inference or during both training and testing, and whether it employs a situation‑based (S‑B) memory mechanism that restricts modulation to a predefined subset of classes (full protocols in Methods, Section~\ref{train_test}). The resulting firing‑rate maps are min–max normalised and passed to a CNN or ViT classifier trained with standard procedures (Methods, Sections ~\ref{sec_Met_CNN}, ~\ref{sec_ViT}). Performance is assessed under ”salt and pepper” impulse noise and partial occlusion noise (Fig. ~\ref{Fig2}C,D; Methods, Section ~\ref{sec_Met_noise}) using classification accuracy and geometric clustering metrics of the learned representations (Methods, Section ~\ref{sec_cluster_metrics}). To improve robustness to occlusions, we applied random scale-and-shift augmentation during training (Fig.~\ref{Fig2}E; Methods, Section~\ref{sec_occl}). Experiments were conducted on three datasets: MNIST, Fashion‑MNIST, and Omniglot. Both MNIST and Fashion-MNIST were binarized prior to training and testing (Methods, Section ~\ref{sec:binarization}). Baselines include the standalone CNN and ViT, as well as a CNN preceded by classical image filters.


\subsection{Learning Performance from Scarce and Noisy Data}\label{subsec2_2}

Modern deep learning models require large labelled datasets and extensive iterative optimisation, which limits their applicability in data‑scarce scenarios. We investigate this limitation by progressively reducing the number of training samples per class. To this end, we performed experiments on two datasets: MNIST and Omniglot. 

MNIST is a standard benchmark consisting of 10 classes of handwritten digits, with 6000 training images and 1000 test images per class. MNIST images were binarized prior to training and testing (Methods, Section ~\ref{sec:binarization}). Binarization is known to reduce the amount of information available to a classifier; this compression inevitably comes at the cost of a notable reduction in classification accuracy compared to using full-precision data \cite{Zhao2021, Qin2020}.

Omniglot is a standard benchmark for few‑shot learning, comprising 1623 character classes with only 20 examples per class, thus explicitly probing the ability of an architecture to learn from very few labelled samples \cite{Lake2015}. In our experiments, for each independent run we randomly sampled a subset of 10 distinct classes from Omniglot, then trained and tested the models on that subset. This procedure was repeated across multiple random subsets to ensure statistically robust performance estimates. The detailed experimental configurations, including the exact training–test splits and the adjustments made to the network architectures for the larger input size, are provided in Methods, Section ~\ref{sec:fashion_omniglot}.

\begin{figure}
\centering
\includegraphics[width=1\textwidth]{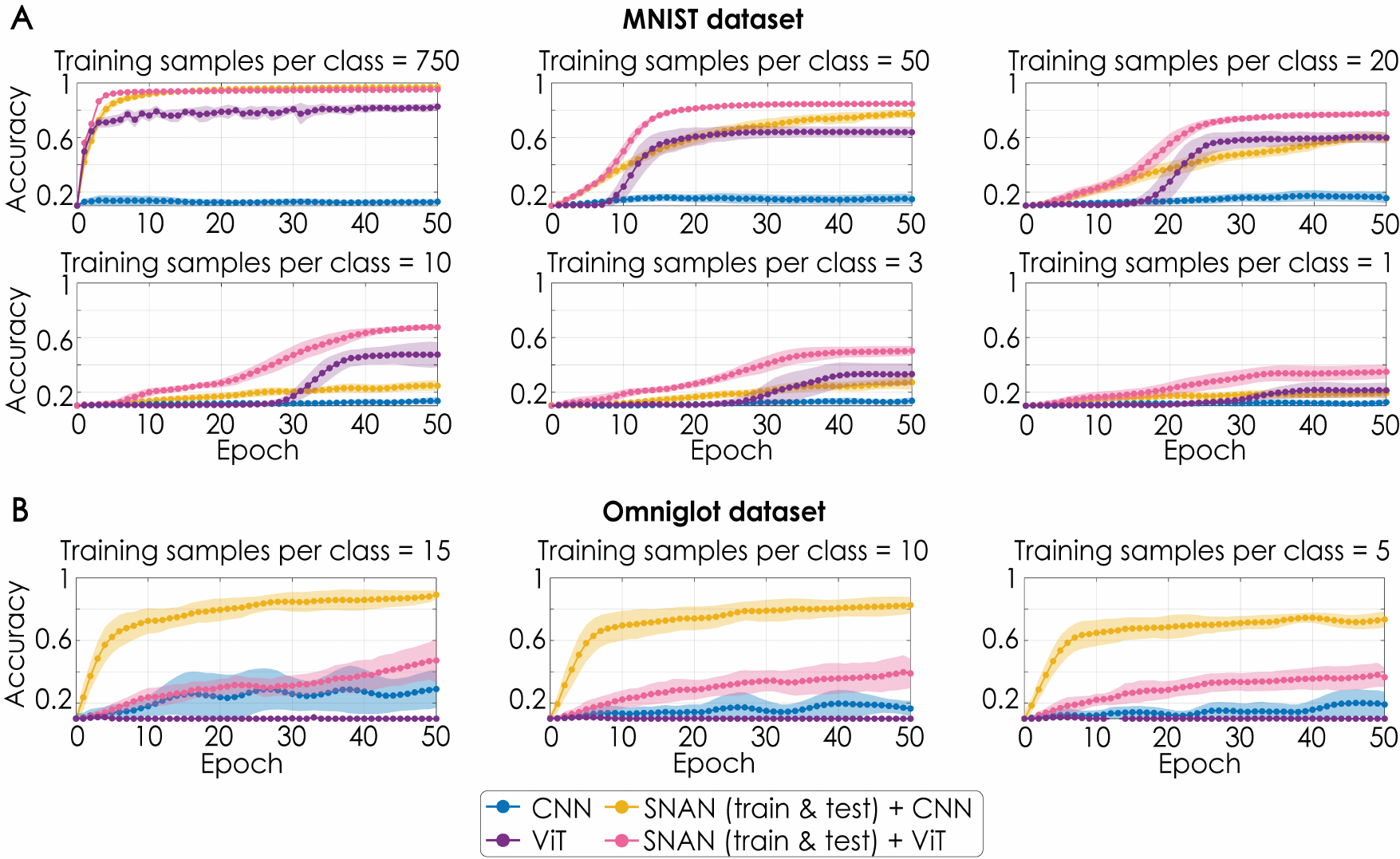}
\caption{Evolution of classification accuracy under data scarcity with test impulse noise (noise level = 0.15). (A) MNIST dataset: training examples per class decreasing from 750 (12.5\% of MNIST) down to 1 (0.017\% of MNIST). (B) Omniglot dataset: training examples per class decreasing from 15 (75\% of Omniglot) down to 5 (25\% of Omniglot). Lines correspond to: SNAN (train and test) + CNN (yellow), standalone CNN (blue), SNAN (train and test) + ViT (pink), standalone ViT (purple).}
\label{Fig3}
\end{figure}

For MNIST dataset, the number of training examples per class was varied from 750 (12.5\% of the full training set) down to a single example (0.017\%) (Fig.~\ref{Fig3}A). For Omniglot, we evaluated the models in three few‑shot settings, using 15, 10, and 5 training examples per class (see Methods, Section~\ref{sec:fashion_omniglot}) (Fig.~\ref{Fig3}B). All architectures were trained for 50 epochs, and classification accuracy was evaluated on a test set after every epoch. Throughout the experiments, the test images were corrupted with impulse noise at a fixed level of 0.15 (see Methods, Section~\ref{sec_rand_noise}).

On MNIST (Fig.~\ref{Fig3}A), the hybrid architectures, SNAN + CNN and SNAN + ViT, exhibit exceptional data efficiency . When trained with 750 samples per class, both hybrids rapidly attain over 90\% accuracy—substantially faster than the standalone ViT, which improves only gradually to $\approx$ 82\%, and the standalone CNN, which never exceeds 20\%. As the training set shrinks, the performance gap widens. With 50 samples per class, the hybrid models still achieve more than 80\% correct classifications, outperforming the standalone ViT ($\approx$ 64\%) by a large margin. Even in the extreme single‑example regime, the neuromorphic hybrids extract usable signal: SNAN + ViT reaches $\approx$ 35\% accuracy, well above the 10\% chance level and markedly better than the standalone ViT ($\approx$ 21\%). The standalone CNN remains below 20\% across all low‑data conditions.

On Omniglot (Fig.~\ref{Fig3}B), the benefits of the neuromorphic preprocessing were even more pronounced. The standalone ViT performed at approximately the chance level (10\%) for all training set sizes, confirming that a modern vision transformer fails to learn meaningful representations from so few examples. The standalone CNN fared slightly better, yet its accuracy remained below 30\%, indicating that even convolutional architectures struggle to generalise under extreme data scarcity. The hybrid SNAN + ViT reached 37-47\% classification accuracy, demonstrating that the SNAN module recovers usable signal from extremely scarce data. The strongest performance was delivered by the hybrid SNAN + CNN, which attained 90\% accuracy when trained with 15 examples per class and maintained a robust 73\% accuracy even when training was limited to just 5 examples per class.

The SNAN component itself learns in a true one‑shot learning: synaptic weights are updated immediately during a single stimulus presentation via spike‑timing‑dependent plasticity (STDP, see Methods, Section ~\ref{sec_Met_SNAN_N}, Eq.~\ref{eq:W_synEE}, ~\ref{eq:W_synIE}), while astrocytic calcium dynamics (Methods, Section ~\ref{sec_Met_SNAN_A}, Eq.~\ref{eq:Ullah}) stabilise the resulting memory traces through neuromodulation (Methods, Section ~\ref{sec_Met_SNAN_BI}, Eq.~\ref{eq:astro_neuro}). This contrasts with the iterative gradient‑based optimisation required by the CNN and ViT.

\subsection{Comparative Analysis of Models in Noisy Images Classification Tasks}\label{subsec2_4}

Having established the data efficiency advantages of the hybrid neuromorphic architectures, we now examine in detail how noise during training and testing affects the performance of different model architectures. We present the analysis in two parts: first for standard random impulse noise, and then for the more challenging case of partial occlusion. Unlike impulse noise, which merely flips individual pixels (see Methods, Section~\ref{sec_rand_noise}, Fig.~\ref{Fig2}C) while leaving the global structure largely intact, partial occlusion introduces structured, semantically meaningful distractors (see Methods, Section~\ref{sec_occl}, Fig.~\ref{Fig2}D). This destroys local features, disrupts spatial context, and forces models to rely on non‑local cues — a demand that conventional CNNs struggle to meet \cite{Zeiler2014}.

We compare four SNAN integration strategies, which differ in whether the SNAN is applied only at inference or during both training and testing, and whether it employs a situation‑based (S‑B) memory mechanism that restricts modulation to a predefined subset of classes (full protocols in Methods, Section~\ref{train_test}). These strategies were tested in combination with both CNN and ViT classifiers. As baselines we evaluate the standalone CNN and ViT, as well as a CNN preceded by classical image filters: mean, median, and bilateral (see Methods, Section~\ref{sec_filters}). For clarity, Fig.~\ref{Fig4} and ~\ref{Fig5} therefore present the most representative outcomes: the standalone CNN and ViT, the four neuromorphic models – SNAN (train~\&~test) + CNN, S‑B~SNAN (train~\&~test) + CNN, SNAN (train~\&~test) + ViT, S‑B~SNAN (train~\&~test) + ViT; and the bilateral filter + CNN combination, as all three filters performed comparably across conditions, with the bilateral filter showing a marginal advantage. The experiments in this section were conducted on binarized MNIST and Fashion‑MNIST, with 750 training images per class. Full results for all architectures are provided in Supplementary Section~\ref{secA}, Fig.~\ref{FigS1}-~\ref{FigS6}.

In addition to classification accuracy, we assessed the quality of the learned representations using four cluster metrics: normalized cluster size $\bar C_s$, mean inter‑cluster distance $\bar C_d$, minimum inter‑cluster distance $\min C_d$, and cluster shift (see Methods, Section~\ref{sec_cluster_metrics} for formal definitions). Briefly, $\bar C_s$ measures the average within‑class dispersion – smaller values indicate tighter, more compact clusters. $\bar C_d$ and $\min C_d$ quantify, respectively, the average and worst‑case separations between class centroids; larger values correspond to better class separation. The cluster shift captures the drift of each class centroid under increasing noise relative to its position on clean data, thereby reflecting the stability of the feature‑space geometry against perturbations.

\subsubsection{Standard Random Noise}

\begin{figure}
\centering
\includegraphics[width=1\textwidth]{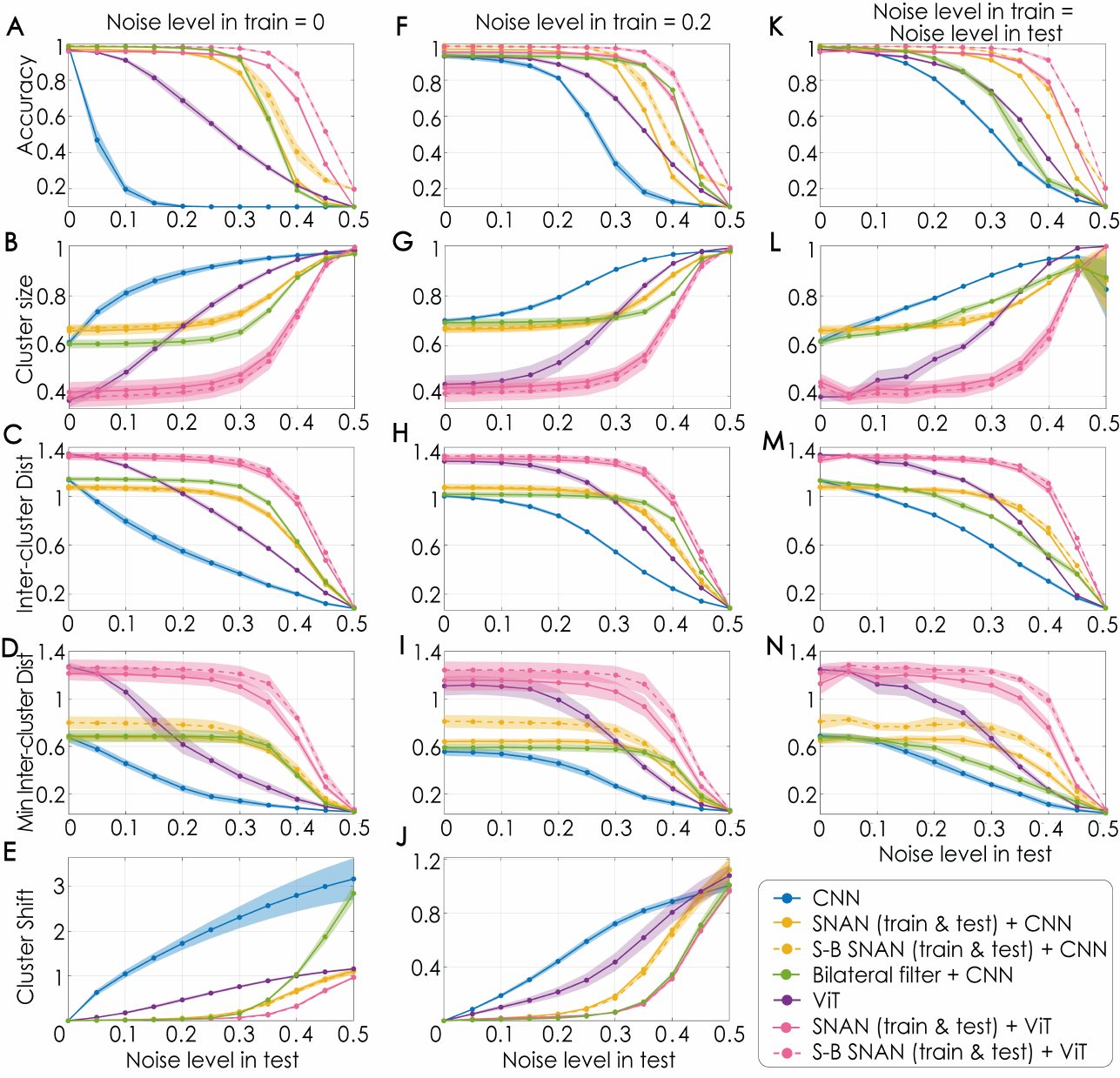}
\caption{Evolution of classification accuracy and feature space properties under increasing impulse noise during testing for models trained under different noise conditions. (A-E) Models trained on clean images. (F-J) Models trained with a fixed noise level ($Noise{\text{ }level} = 0.2$). (K-N) Models trained with noise level matched to the test condition. (A, F, K) Classification accuracy. (B, G, L) Normalized cluster size of learned representations, $\bar{C_s}$. (C, H, M) Pairwise inter-cluster distance, $\bar{C_d}$. (D, I, N) Nearest-class inter-cluster distance. (E, J) Cluster shift. All results are averaged over 10 runs; the solid lines represent the mean and the shaded regions correspond to $\pm$ SD.}
\label{Fig4}
\end{figure}

On the MNIST dataset, Fig.~\ref{Fig4} summarises the evolution of classification accuracy and feature‑space geometry for all main architectures—CNN, ViT, their hybrids with SNAN (standard and situation‑based), and the best‑performing classical filter (bilateral filter + CNN)—under increasing test‑time impulse noise. Three training regimes are compared: clean images only, a fixed noise level of 0.2, and noise matched to the test condition. Full results for all tested models, including every filter type, are provided in Supplementary Section~\ref{secA}, Figs.~\ref{FigS1} and~\ref{FigS2}. The Fashion-MNIST dataset confirms a similar pattern of results; the corresponding figure is provided in Supplementary Fig.~\ref{FigS5}.

\textbf{Training on Clean Images}.
In the absence of noise in test images, all architectures achieve high classification accuracy above 96\% (Fig.~\ref{Fig4}A-E). As test-time impulse noise increases, however, performance diverges sharply. The standalone CNN drops to chance level at noise $\ge 0.15$, whereas the standalone ViT degrades to $\approx 69\%$ at noise $0.2$ and to $\approx 43\%$ at $0.3$. The most noise-robust architecture is S-B SNAN (train \& test) + ViT, which maintains $>95\%$ accuracy across noise levels from $0$ to $0.35$ inclusive. The non-situational SNAN (train \& test) + ViT retains $\approx 93\%$ at noise $\le 0.3$ and $\approx 88\%$ at $0.35$. Among the CNN-based hybrids, SNAN (train \& test) + CNN and S-B SNAN (train \& test) + CNN both keep accuracy above $80\%$ up to noise level $0.3$. The classical filter baselines (bilateral filter + CNN) perform comparably to CNN-based hybrids for noise $\le 0.35$, but are consistently outperformed by the ViT-based hybrids.

Geometric analysis of the learned feature spaces corroborates these accuracy trends. As noise increases, the normalized cluster size $\bar{C}_s$ expands (Fig.~\ref{Fig4}B) while both the mean inter-cluster distance $\bar{C}_d$ (Fig.~\ref{Fig4}C) and the minimum inter-cluster distance (Fig.~\ref{Fig4}D) contract, indicating progressive feature dispersion and class-boundary collapse. At a representative noise level of $0.3$, the standalone CNN shows near-complete breakdown of cluster structure, consistent with its chance-level accuracy. The standalone ViT exhibits substantial dispersion ($\bar{C}_s \approx 0.84$) and collapsing separation ($\bar{C}_d \approx 0.73$, $\min[\bar{C}_d] \approx 0.35$; Fig.~\ref{Fig4}B-D). In contrast, the S-B SNAN (train \& test) + ViT hybrid maintains compact and well-separated clusters ($\bar{C}_s \approx 0.46$, $\bar{C}_d \approx 1.29$, $\min[\bar{C}_d] \approx 1.21$; Fig.~\ref{Fig4}B-D). The non-situational SNAN + ViT also preserves markedly better geometry than the standalone ViT. Among CNN-based hybrids, SNAN (train \& test) + CNN and S-B SNAN (train \& test) + CNN achieve intermediate cluster metrics ($\bar{C}_s \approx 0.73$, $\bar{C}_d \approx 0.98$, $\min[\bar{C}_d] \approx 0.72$). The bilateral filter + CNN baseline performs slightly worse than the CNN-based neuromorphic hybrids, confirming that classical filtering does not fully replicate the SNAN's geometrically stabilizing effect. Cluster centroid shifts (Fig.~\ref{Fig4}E) further confirm that hybrid models experience much smaller displacements than their standalone counterparts.

\textbf{Training with Fixed Noise}. 
Next, we add medium noise (noise level 0.2) to the training data (Fig.~\ref{Fig4}F–J). Adding noise during training improves all models, yet the hierarchy remains unchanged. The standalone CNN and ViT gain accuracy but continue to lag behind their hybrid counterparts. Among CNN-based hybrids, S-B SNAN (train \& test) + CNN achieves $>93\%$ accuracy up to noise level $0.3$, while SNAN (train \& test) + CNN reaches $87\%$ (Fig.~\ref{Fig4}F). The ViT-based hybrids again set the upper bound: S-B SNAN + ViT sustains approximately $96\%$ accuracy at noise $0.35$, and the standard SNAN + ViT attains approximately $90\%$. At higher noise levels ($>0.3$), the bilateral filter + CNN pipeline outperforms the CNN-based hybrids (e.g., $88\%$ vs.\ $\approx 78\%$ for S-B SNAN + CNN at noise level $0.35$), but remains well below the ViT-based hybrids, which keep $>90\%$. Geometrically, classical filtering and the SNAN + ViT hybrids yield the smallest centroid displacement at high noise, followed by the CNN-based hybrid models, while standalone architectures exhibit the largest shifts (Fig.~\ref{Fig4}J). This indicates that training with a moderate, fixed noise level already stabilises the feature space, and neuromorphic preprocessing provides a clear advantage over purely baseline models.

\textbf{Matched-noise Training}.
Now we train with noise levels matched to the test condition (Fig.~\ref{Fig4}K-N). When the training noise level is matched to the test condition, the advantage of SNAN preprocessing becomes decisive. The ViT‑based hybrids set a new performance ceiling: S‑B SNAN + ViT reaches $97\%$ accuracy, and the standard SNAN + ViT attains $90\%$ at noise $0.35$ (Fig.~\ref{Fig4}K). Remarkably, even at noise $0.4$, where all CNN‑based models fall below $60\%$, S‑B SNAN + ViT retains $91\%$ accuracy. Among CNN‑based hybrids, S‑B SNAN (train \& test) + CNN achieves $91\%$ at the same noise level, outperforming SNAN (train \& test) + CNN ($82\%$) and the standalone ViT ($58\%$). Geometric metrics echo these findings: at noise $0.35$, the S‑B SNAN + ViT variant maintains the smallest cluster size ($\bar{C}_s \approx 0.51$), the largest mean inter‑class separation ($\bar{C}_d \approx 1.25$), and the widest safety margin ($\min[\bar{C}_d] \approx 1.16$) among all tested configurations (Fig.~\ref{Fig4}L-N).

In all regimes, the SNAN preprocessing layer universally enhances the robustness of downstream classifiers, with the situation‑based variant and noise‑matched training giving the strongest gains. The results point to a robust, biologically inspired filtering mechanism that cannot be replicated by classical image denoising. Equivalent experiments on the Fashion‑MNIST dataset confirm the same pattern of results (Supplementary Section~\ref{secA}, Fig.~\ref{FigS5}).

\subsubsection{Partial Occlusion}\label{subsec2_4_2}

Having established robustness to impulse noise, we now evaluate a more complex corruption type: partial occlusion (Fig.~\ref{Fig2}D), which poses greater challenges due to semantic interference -- the occluding content itself forms a valid image pattern. In this setting, a randomly chosen digit pattern is overlaid onto the target image, covering a specified \textit{occlusion size} (percentage of the target image area; see Methods, Section~\ref{sec_occl}). All models were trained with random scale‑and‑shift augmentation (Fig.~\ref{Fig2}E; Methods, Section~\ref{sec_occl}), meaning that the original binarized MNIST images were randomly scaled and translated within given ranges. This deliberately increases the difficulty of the training task while encouraging invariance to occlusions.

\begin{figure}
\centering
\includegraphics[width=1\textwidth]{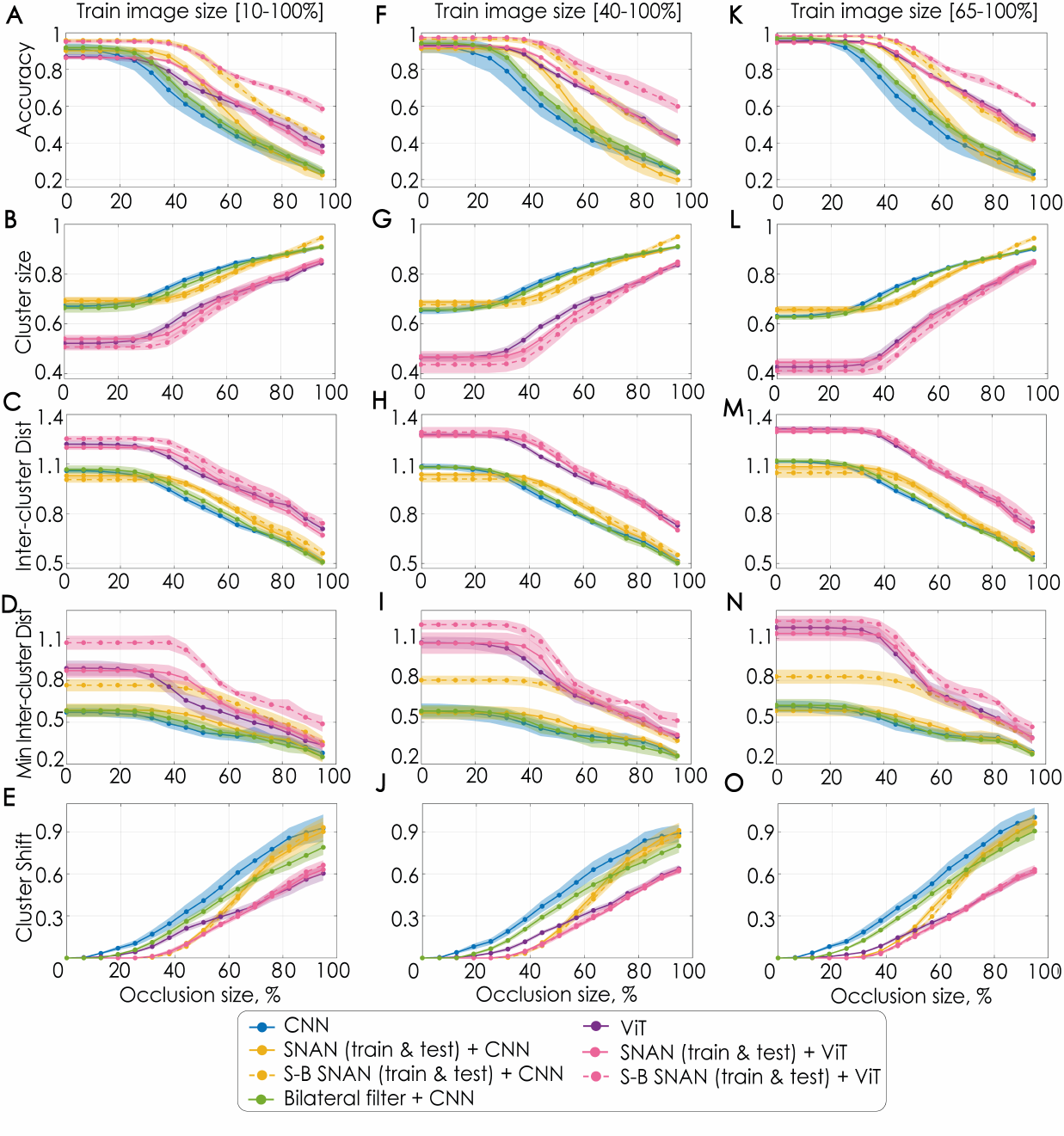}
\caption{Evolution of classification accuracy and feature space properties with increasing occlusion size (percentage of target image area) for models trained using different scale augmentation ranges. (A-E) Trained with scale augmentation $X,Y \sim \mathcal{U}$[10\%,100\%], (F-J) Trained with scale augmentation $X,Y \sim \mathcal{U}$[40\%,100\%], (K-O) Trained with scale augmentation $X,Y \sim \mathcal{U}$[65\%,100\%]. (A,F,K) Classification accuracy, (B,G,L) Normalized cluster size of learned representations, $\bar{C_s}$. (C, H, M) Pairwise inter-cluster distance, $\bar{C_d}$. (D, I, N) Nearest-class inter-cluster distance. (E, J, O) Cluster shift. All results are averaged over 10 runs; the solid lines represent the mean and the shaded regions correspond to $\pm$ SD.}
\label{Fig5}
\end{figure}

Fig.~\ref{Fig5} reports accuracy and feature‑space geometry for the main architectures -- CNN, ViT, their SNAN hybrids (standard and situation‑based), and the best classical filter (bilateral filter + CNN) -- under increasing occlusion size. Three augmentation ranges are compared: broad ($X,Y \sim \mathcal{U}[10\%,100\%]$), moderate ($[40\%,100\%]$), and narrow ($[65\%,100\%]$). Complete curves for all tested variants, including every filter, are provided in Supplementary Section~\ref{secA}, Figs.~\ref{FigS4} and~\ref{FigS5}.

\textbf{Broad Augmentation.}
Under the most diverse augmentation regime, the situation‑based SNAN variants consistently outperform all alternatives (Fig.~\ref{Fig5}A-E). At $50\%$ occlusion, S‑B SNAN (train \& test) + ViT reaches $87\%$ accuracy, surpassing the standard SNAN + ViT ($74\%$) and the standalone ViT ($68\%$). Among CNN‑based hybrids, S‑B SNAN + CNN achieves $87\%$, while the non‑situational SNAN + CNN reaches $72\%$; the standalone CNN attains only $55\%$. The bilateral filter + CNN pipeline offers no meaningful gain, as it cannot semantically distinguish the occluding digit from the target. Geometric analysis reinforces these findings: at $50\%$ occlusion, S‑B SNAN + ViT preserves more compact clusters ($\bar{C}_s \approx 0.62$) and a substantially larger minimum inter‑class margin ($\min[\bar{C}_d] \approx 0.91$) than the standalone ViT ($\bar{C}_s \approx 0.67$, $\min[\bar{C}_d] \approx 0.61$).

\textbf{Moderate and Narrow Augmentation.}
When the augmentation range is narrowed to $[40\%,100\%]$, all models improve, yet hybrids retain a clear edge. At $50\%$ occlusion, S‑B SNAN + ViT achieves $90\%$ accuracy and S‑B SNAN + CNN $86\%$, while the standalone ViT reaches $77\%$ and the standalone CNN $54\%$. Under the narrowest augmentation ($[65\%,100\%]$), S‑B SNAN + ViT attains $92\%$ accuracy -- the highest recorded across all occlusion experiments -- and S‑B SNAN + CNN reaches $89\%$; the standalone ViT achieves $82\%$ but degrades more rapidly at larger occlusions (Fig.~\ref{Fig5}K-O). Across all regimes, geometric metrics confirm that hybrid models maintain more compact, better‑separated feature spaces than their standalone counterparts. Full results for all tested architectures are provided in Supplementary Figs.~\ref{FigS3} and~\ref{FigS4}.

In all occlusion experiments, the SNAN preprocessing layer provides a universal mechanism for robustness that effectively complements both CNN and ViT classifiers. Its advantage cannot be replicated by classical filters, because astrocytic calcium dynamics selectively suppress irrelevant occlusion content rather than operating on fixed local image statistics. The situation‑based variant, in particular, delivers the best overall performance. Equivalent experiments on the Fashion‑MNIST dataset confirm the same pattern of results (Supplementary Section~\ref{secA}, Fig.~\ref{FigS6}).

\section{Discussion}\label{sec_discussion}
 
The central finding of this work is that incorporating biologically motivated spiking neuron–astrocyte dynamics as a preprocessing layer into conventional deep learning architectures produces qualitative transformation. Even though several recent studies demonstrated the viability of embedding astrocytes into conventional neural network models \cite{Kozachkov2025,Kozachkov2023}, a systematic study of performance gains due to these additions relative to classical architectures has not been done. Here we, for the first time, show that neuromorphic-inspired architectural modifications give rise to significant levels of robustness to noise and data efficiency that cannot be achieved by architectural scaling or training strategies alone.These new hybrid models can learn from extremely scarce data. They also show significant robustness to noise. These performance characteristics could not be achieved within standard architectures. Several aspects of these results merit further discussion. 

A striking pattern emerges when architectures are ordered by their proximity to biological neural computation: standard feed‑forward networks → CNNs → vision transformers → SNAN‑based hybrids. Performance degradation under noise and data scarcity decreases precisely along this ordering: the closer an architecture is to its biological prototype, the greater its resilience. We hypothesise that this is not coincidental but reflects a deeper principle: each architectural step toward biological computation introduces inductive biases that are well matched to the statistical structure of natural sensory signals. CNNs exploit spatial locality, transformers capture long‑range dependencies through attention, and the SNAN layer adds temporal multi‑scale processing and context‑dependent gating—mechanisms that biological systems have evolved specifically for robust perception in noisy, information‑scarce environments.

The mechanism underpinning the SNAN’s effectiveness appears to be twofold. First, astrocytic calcium dynamics implement a form of adaptive, spatially coherent noise suppression that differs fundamentally from conventional denoising filters. While median and bilateral filters operate on fixed local statistics, the SNAN’s filtering is stimulus‑dependent: astrocytes selectively amplify coherent neural activity patterns that correspond to learned stimuli while suppressing incoherent activations caused by noise. This explains why SNAN‑based hybrids dramatically outperform classical filter‑plus‑CNN pipelines under partial occlusion—a corruption type where conventional filters are ineffective because the occluding content is itself a valid image pattern rather than statistical noise. Second, the dual‑timescale architecture supports a form of rapid contextual memory that allows the network to maintain a “situational prior” during inference, effectively constraining the hypothesis space and reducing the data requirements for reliable classification.

Beyond the incorporation of astrocytic dynamics, recent efforts to advance data processing systems have explored additional neurobiological mechanisms to overcome the limitations of conventional deep learning. For instance, a major challenge for standard models is catastrophic forgetting during continual learning, where exposure to a new task leads to the loss of previously acquired knowledge. Establishing lifelong learning in dynamically changing environments is a nascent but critical direction, and it remains an open question whether the additional temporal dimension of SNNs could facilitate such capabilities \cite{Zuo2017}. Closely related is the challenge of learning with fewer data; unsupervised learning in SNNs, combined with minimal supervision, offers a promising route for data-efficient training \cite{Kheradpisheh2018, Masquelier2007}. Drawing direct inspiration from neuroscience has yielded other efficient strategies: Masquelier et al. \cite{Masquelier2007} demonstrated that STDP with temporal coding can produce a convolutional hierarchy of feature-selective neurons, mimicking the visual cortex pathway. Furthermore, incorporating dendritic learning \cite{Rao2001} and structural plasticity \cite{Roy2017} presents interesting possibilities for improving spike-based learning. Finally, liquid state machines (LSMs)\cite{Maass2011}, which pair unstructured recurrent spiking networks with a simple linear readout, have shown success in sequential recognition tasks \cite{Schrauwen2008}.

It is important to acknowledge the scope and limitations of the present study. All experiments were conducted on three widely used datasets—MNIST, Fashion‑MNIST, and Omniglot. Although canonical, these datasets are low‑dimensional and grayscale; moreover, all images were binarised (reduced to black and white), a step that has the potential to restrict information available in the data and lower the baseline accuracy even on clean images. The fact that the hybrid models retain a substantial advantage under such  conditions indicates that the identified benefits are fundamental rather than dataset‑specific. Nevertheless, we do not claim that the observed performance advantages will transfer without modification to more complex, higher‑dimensional tasks such as natural image classification with colour channels. Extending the SNAN to colour images would substantially increase the complexity of the neuromorphic layer. We also note that our ViT baseline, while substantially more expressive than the CNN baseline, remains a relatively modest architecture compared with the large‑scale vision transformers used in practice; the interaction between SNAN preprocessing and larger‑capacity classifiers remains an open question. These considerations define a clear pathway for future work: systematic evaluation on increasingly complex benchmarks, development of efficient SNAN implementations on neuromorphic hardware, and investigation of the scalability of the proposed hybrid framework.
 
\section{Conclusion}\label{sec_conclusion}
 
We introduced a novel class of hybrid neuromorphic models that integrate Spiking Neuron–Astrocyte Networks (SNANs) with both convolutional neural networks and vision transformers. Through systematic experiments on three datasets -- MNIST, Fashion‑MNIST, and Omniglot—under conditions of noise corruption and data scarcity, we demonstrated that these hybrids achieve levels of performance that are hardly unattainable by their conventional counterparts of comparable size but without neuromorphic additions. The SNAN preprocessing layer provides intrinsic noise filtering and contextual memory formation through biologically motivated dual‑timescale dynamics, universally enhancing any downstream classifier. The best‑performing new hybrid configurations,  maintain high accuracy under matched noise training at corruption levels where standalone architectures collapse. Moreover, the new models maintain their generalization capabilities when training data are scarce -- the quality that none of the standard models were able to replicate. All images were binarised, further underscoring the robustness of the proposed approach.

We term this phenomenon neuromorphic supremacy, by analogy with quantum supremacy: just as quantum circuits solve certain problems inaccessible to classical computers, neuromorphic architectures equipped with astrocytic modulation and spiking dynamics solve perception tasks that remain fundamentally challenging for conventional models and algorithms of deep learning. 


\bmhead{Code availability} 
The code is available at https://github.com/altergot/Hybrid\_\linebreak Neuromorphic\_Models\_for\_image\_classification.

\bmhead{Acknowledgements}

\bmhead{Declarations}
The authors declare that they have no known competing financial interests or personal relationships that could have appeared to influence the work reported in this paper.

\section{Methods}\label{sec_Met}

\subsection{Spiking Neuron-Astrocyte Network}\label{sec_Met_SNAN}

\begin{figure}
\centering
\includegraphics[width=1\textwidth]{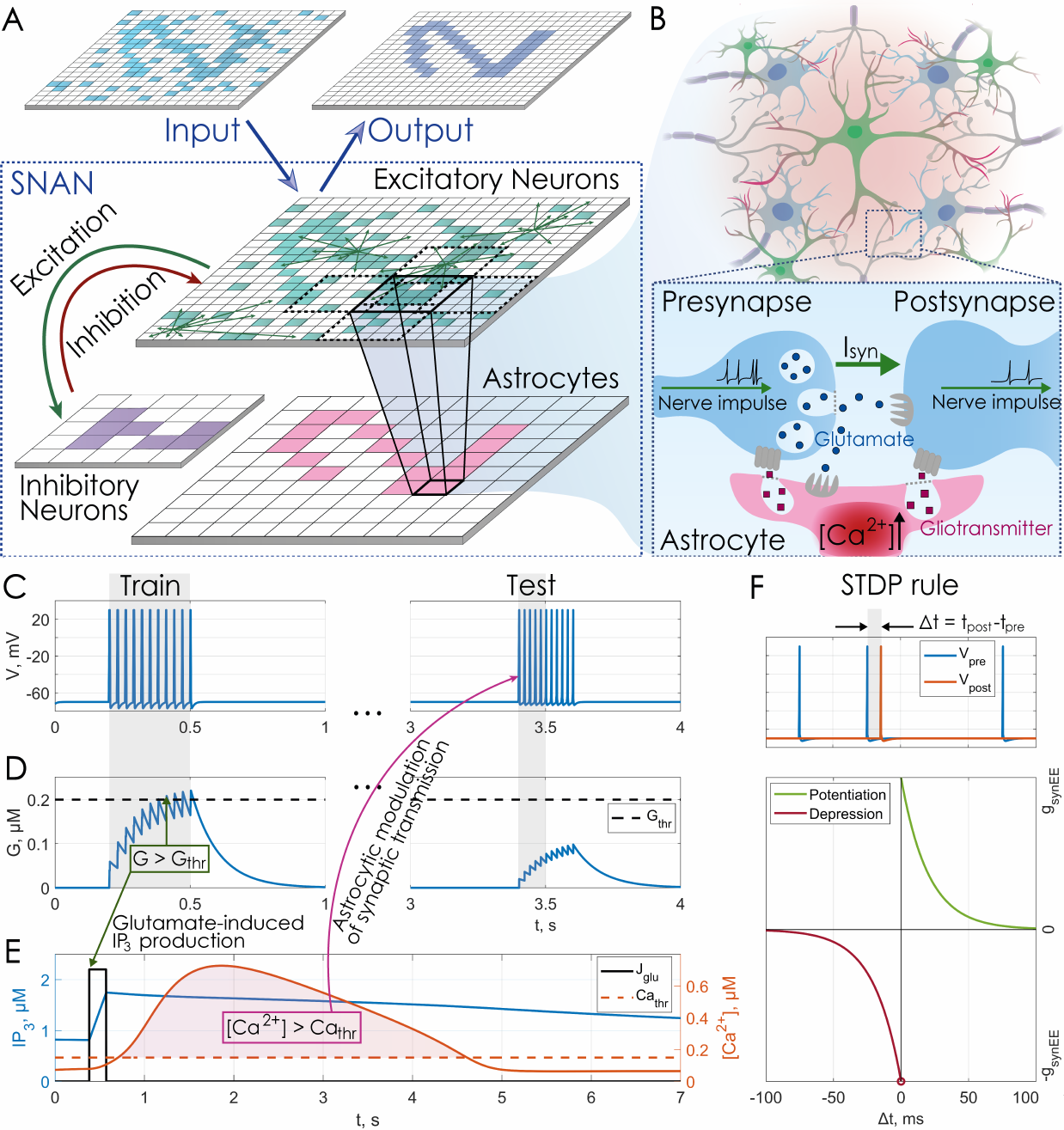}
\caption{ A Spiking neuron-astrocyte network (SNAN) model overview. (A) - SNAN model topology. (B) - Schematic illustration of bidirectional neuron-astrocyte interactions. (C) - Spike train and (D) concentration of the neurotransmitter, $G(t)$, of the stimulus-specific neuron. (E) Intracellular concentration of $Ca^{2+}$ and $IP_3$ in a stimulus-specific astrocyte. Black bars at the top indicate periods when training and test stimuli were presented. In response to the presynaptic spike train (C), the neurotransmitter, glutamate, $G$ releases (D) into extracellular space and the concentration of $IP_3$ increases in the astrocyte (E, blue line) inducing the elevation of intracellular $Ca^{2+}$ (E, red line). (F) - The STDP rule updates synaptic weights according to the timing difference between pre- and postsynaptic spikes.}
\label{FigM1}
\end{figure}

The concept and architecture of the SNAN model are schematically summarized in Fig.~\ref{FigM1}, and all relevant parameters of the model are provided in Table \ref{tab:SNAN_params}.  This biologically motivated computational model implements short-term memory through interactions between neural and astrocytic networks, operating across distinct timescales: the millisecond-scale dynamics of spiking neurons and the second-scale dynamics of astrocytic calcium signaling \cite{Gordleeva2023}. The neuronal network consists of sparsely connected excitatory and inhibitory spiking neurons with plastic synapses, trained using a traditional spike-timing-dependent plasticity (STDP) rule. Astrocytes monitor neural activity and respond through intracellular calcium elevations, inducing short-term synaptic plasticity that generates localized spatial synchronization in neuronal ensembles. The resulting astrocyte-mediated short-term memory exhibits one-shot learning capability and maintains information persistence. Astrocytic modulation implements a Hebbian-like plasticity mechanism that selectively strengthens coherent neural activations while filtering nonspecific activity. The architecture synergistically combines fast spiking neurons for information readout with slow astrocytic processes for storage, enabling efficient memory loading. The SNAN architecture features three functionally interconnected layers: pyramidal neurons, interneurons, and astrocytes. Input signals encoded as 2D patterns project onto pyramidal neurons, which establish random connections with exponentially distributed lengths. Bidirectional connectivity between pyramidal neurons and interneurons maintains excitation-inhibition balance. Astrocytes form gap-junction-coupled networks that detect coherent pyramidal firing through neurotransmitter sensing. When activated, astrocytes generate calcium signals that trigger gliotransmitter release, modulating synaptic weights in the corresponding neuronal groups. The system's output is derived from the transient discharge frequencies of pyramidal neurons.

In this section, the SNAN architecture is described in detail together with the STDP learning rule and neuron/astrocytic models. Specifically, we start with biologically realistic models of neuronal, astrocytic networks that capture the essence of the biological interplay between these cells, at the same time
minimizing the computational overhead. Then, we describe the communication between pyramidal neurons and astrocytes at tripartite synapses.

\subsubsection{Neural Network}\label{sec_Met_SNAN_N}
The dynamics of neuronal membrane potential is given by \cite{Izhikevich2003}:
\begin{equation}
\label{eq:Izh_main}
\begin{aligned}
&\frac{dV^{(i,j)}}{dt} = 0.04{V^{(i,j)}}^{2} + 5V^{(i,j)} - U^{(i,j)} + 140 +I_{\text{app}}^{(i,j)} + I_{\text{syn}}^{(i,j)}, \\
&\frac{dU^{(i,j)}}{dt} = a ( b V^{(i,j)} - U^{(i,j)});\\
\end{aligned}
\end{equation}

\noindent with initial conditions $V^{(i,j)}(0)$ = -65 mV,\quad $U^{(i,j)}(0)$ = $b \cdot V^{(i,j)}(0)$. The auxiliary after-spike resetting is: 

 
\begin{equation}
    \label{eq:Izh_cond}
    \text{if } V^{(i,j)} \ge \text{30 mV, then}
        \begin{cases}
            V^{(i,j)}\gets c \\U^{(i,j)} \gets U^{(i,j)} + d.\\
        \end{cases}
\end{equation}
 
The superscript $(i,j)$ denotes the neural index in the network. $I_{\text{app}}^{(i,j)}$  is the input current, derived from the digital input image and applied only to excitatory neurons. The total synaptic current $I_{\text{syn}}^{(i,j)}$ received from all presynaptic neurons is described by \cite{Kazantsev2011, Esir2018}:
 
\begin{equation}
    \label{eq:I_syn_EE_IE}
    \begin{aligned}
    &I_{\text{syn}}^{(i,j)} =\sum_{k=1}^{N_{in}^{(i,j)}} \frac{w_{syn,k} (E_{\text{syn}} -V^{(i,j)})} {1+\exp(-V_{\text{pre,k}}/k_{\text{syn}})},\\
    \end{aligned}
\end{equation}
 
\noindent where $N_{in}^{(i,j)}$ is the total number of synapses, $w_{syn,k}$ is the weight of the $k$-th synapse connected to neuron $(i,j)$, $V_{pre}$ is the presynaptic neuron's membrane potential, $E_{syn}$ = 0 mV for excitatory synapses and $E_{syn}$ = -90 mV for inhibitory synapses.
 
The synaptic connection architecture between neurons is non-specific (random) with an exponential distribution of connection lengths. The number of output connections per neuron is fixed at $N_{out}$. Postsynaptic neuron coordinates are calculated in a polar coordinate system with distance $r$ following the exponential distribution:
\begin{equation}
f_R(r) = 
\begin{cases} 
\frac{1}{\lambda_{syn}} \exp(-r/\lambda_{syn}), & r \geq 0 \\
0, & r < 0
\end{cases}
\label{eq:radius_dist}
\end{equation}
and angle $\phi$ uniformly distributed in $[0, 2\pi]$. Coordinates are computed with layer scaling:
\begin{align}
&\text{EE: }  &x_{post} &= \left\lceil x_{pre} + r \cos\phi \right\rceil,  &y_{post} &= \left\lceil y_{pre} + r \sin\phi \right\rceil \\
&\text{EI: }  &x_{post} &= \left\lceil K_1^{-1}x_{pre} + r \cos\phi \right\rceil,  &y_{post} &= \left\lceil K_2^{-1}y_{pre} + r \sin\phi \right\rceil \\
&\text{IE: }  &x_{post} &= \left\lceil K_1 x_{pre} + r \cos\phi \right\rceil,  &y_{post} &= \left\lceil K_2 y_{pre} + r \sin\phi \right\rceil
\label{eq:coord_transform}
\end{align}
where $K_1 = W/W_1$, $K_2 = H/H_1$ (scaling coefficients). Duplicate connections are excluded by re-sampling.
 
Excitatory neurons interact with each other (EE connections) and with inhibitory neurons (EI connections). Inhibitory neurons connect to excitatory neurons (IE connections) and are not interconnected. The architecture of synaptic connections between neurons is non-specific with different parameters within excitatory and inhibitory layers.
 
The synaptic weights dynamically change during training only for EE and IE synaptic connections (synaptic weights for EI synapses are fixed). The STDP rule updates synaptic weights according to the timing difference between pre- and postsynaptic spikes:
 
\begin{equation}
    \label{eq:W_synEE}
    \begin{aligned}
    \delta w_{synEE,k}(\Delta t) = \begin{cases}
                        g_{synEE}\exp(\Delta t/\tau),  \ \Delta t \geq 0 ,\\
                       -g_{synEE}\exp(\Delta t/\tau),  \ \Delta t < 0 ;\\
                         \end{cases}\\
    w_{synEE,k} \in [10^{-4}, w_{synEEmax}],
    \end{aligned}
\end{equation}
 
\noindent where $\delta w_{synEE,k}(\Delta t)$ is the synaptic weight update, $\Delta t$ is the time difference between post- and presynaptic spikes, $g_{synEE}$ is the plasticity window height, and $\tau$ controls the plasticity window width.   
 
Training of IE synaptic connections is designed such that inhibitory neurons activated by excitatory neurons inhibit the entire non-stimulus-specific subset of excitatory neurons. Therefore, IE synapse weights are updated according to:
 
\begin{equation}
    \label{eq:W_synIE}
    \begin{aligned}
    \delta w_{synIE,k}(\Delta t) = \begin{cases}
                        g_{synIE}\exp(\Delta t/\tau) \Theta(f^*-f),  \ \Delta t \geq 0 ,\\
                       -g_{synIE}\exp(\Delta t/\tau),  \ \Delta t < 0 ;\\
                         \end{cases}\\
    w_{synIE,k} \in [10^{-4}, w_{synIEmax}],
    \end{aligned}
\end{equation}
 
\noindent where $\Delta t$ is the time difference between post- and presynaptic spikes, $g_{synIE}$ is the plasticity window height, and $\tau$ controls the plasticity window width. Variables $f$ and $f^*$ represent the firing rate of the postsynaptic excitatory neuron and the threshold firing rate, respectively. $\Theta$ is the Heaviside step function.
 
\subsubsection{Astrocytic Network}\label{sec_Met_SNAN_A}
 
The Ullah model \cite{ULLAH2006} is used to describe the dynamics of the intracellular concentrations of IP$_3$ and Ca$^{2+}$ in astrocytes:
 
\begin{equation}
\label{eq:Ullah}
\begin{aligned}
&\frac{d[IP_3]^{(m,n)}}{dt}  = \frac{[IP_3]^*-[IP_3]^{(m,n)}}{\tau_{IP_3}}+J_{\text{PLC$\delta$}}^{(m,n)}+J_{\text{glu}}^{(m,n)}+diff_\text{IP3}^{(m,n)},\\
&\frac{d[Ca^{2+}]^{(m,n)}}{dt}  = J_{\text{ER}}^{(m,n)}-J_{\text{pump}}^{(m,n)}+J_{\text{leak}}^{(m,n)}+J_{\text{in}}^{(m,n)}-J_{\text{out}}^{(m,n)}+diff_\text{Ca}^{(m,n)},\\
&\frac{dh^{(m,n)}}{dt}  = a_2 \left(d_2\frac{[IP_3]^{(m,n)}+d_1}{[IP_3]^{(m,n)}+d_3}(1-h^{(m,n)})-[Ca^{2+}]^{(m,n)} h^{(m,n)} \right);\\
\end{aligned}
\end{equation} 
 
\noindent with initial conditions $[IP_3]^{(m,n)}(0) = 0.8202\ \mu\text{M}$, $[Ca^{2+}]^{(m,n)}(0) = 0.0725\ \mu\text{M}$, 
$h^{(m,n)}(0) = 0.8863$. The superscript $(m,n)$ denotes the astrocyte index in the network, $h$ represents the fraction of open $IP_3$-dependent calcium receptors. $J_{\text{$ER$}}$, $J_{\text{$pump$}}$, $J_{\text{$leak$}}$, $J_{\text{$in$}}$ and $J_{\text{$out$}}$ are Ca$^{2+}$ fluxes. Currents $diff_\text{Ca}^{(m,n)}$ and $diff_\text{IP3}^{(m,n)}$ describe the diffusion of IP$_3$ molecules and Ca$^{2+}$ ions through gap junctions from nearby astrocytes \cite{Yamamoto1990,Nagy2000}. $J_{\text{$glu$}}$ describes the glutamate-induced IP$_3$ production in response to neural activity.
 
The calcium and IP$_3$ fluxes in the Ullah model (\ref{eq:Ullah}) are defined as:
\begin{align}
&J_{\text{ER}} = c_1v_1[Ca^{2+}]^3 h^3[IP_3]^3 \frac{c_0/c_1 - (1+1/c_1)[Ca^{2+}]}{([IP_3]+d_1)^3 ([Ca^{2+}]+d_5)^3} \\
&J_{\text{pump}} = \frac{v_3[Ca^{2+}]^2}{k_3^2 + [Ca^{2+}]^2} \\
&J_{\text{leak}} = c_1v_2(c_0/c_1 - (1+1/c_1)[Ca^{2+}]) \\
&J_{\text{in}} = \frac{v_6[IP_3]^2}{k_2^2 + [IP_3]^2} \\
&J_{\text{out}} = k_1[Ca^{2+}] \\
&J_{\text{PLC}\delta} = \frac{v_4([Ca^{2+}] + (1-\alpha)k_4)}{[Ca^{2+}] + k_4}
\label{eq:ullah_fluxes}
\end{align}
Diffusion through gap junctions:
\begin{align}
&diff_\text{Ca}^{(m,n)} = d_{ca}\sum_j ([Ca^{2+}]_j - [Ca^{2+}]^{(m,n)}) \\
&diff_\text{IP3}^{(m,n)} = d_{ip3}\sum_j ([IP_3]_j - [IP_3]^{(m,n)})
\label{eq:gap_junctions}
\end{align}
where $j$ denotes neighboring astrocytes.
 
\subsubsection{Bidirectional Neuron-Astrocyte Interaction}\label{sec_Met_SNAN_BI}
 
Each astrocyte in the SNAN bidirectionally interacts with ensemble of $N_{AE}$ ($L\times L$ dimension) excitatory neurons with some overlapping \cite{Halassa2007} (Fig.~\ref{FigM1} B). Spiking neuronal activity (Fig.~\ref{FigM1} C) induces the release of neurotransmitter (glutamate) from the presynaptic terminals (Fig.~\ref{FigM1} D) and triggers the production of inositol 1,4,5-trisphosphate (IP$_3$) in astrocytes, which is followed by the generation of a calcium pulse (Fig.~\ref{FigM1} E).
 
The dynamics of extracellular glutamate concentration released by the $(i,j)$th neuron into the synaptic cleft is described by \cite{Gordleeva2012,Pankratova2019}:
 
\begin{equation}
\label{eq:Glu}
\begin{aligned}
\frac{d[G]^{(i,j)}}{dt} & = -\alpha_\text{glu} [G]^{(i,j)}+k_\text{glu}\Theta\left(V^{(i,j)}-30 mV\right),\\
\end{aligned}
\end{equation}
 
\noindent with initial conditions $[G]^{(i,j)}(0) = 0\ \mu\text{M}$. $\Theta$ is the Heaviside step function, $\alpha_{glu}$ is the glutamate clearance rate constant, and $k_{glu}$ is the release efficacy.
Glutamate-induced $IP_3$ production is modeled as a rectangular pulse with amplitude $A_{glu}$ and duration $t_{glu}$:
 
\begin{equation}
    \label{eq:Jglu}
    \begin{aligned}
    J_{\text{glu}} = \begin{cases}
                        A_\text{glu}, \qquad &\text{if} \quad t_0<t\le t_0+t_{\text{glu}},  \\
                        0, \qquad &\text{otherwise};\\
                     \end{cases}
    \end{aligned}
\end{equation}
 
The time point $t_0$ corresponds to the time when the extracellular concentration of released glutamate of at least $F_{act}\cdot N_{AE}$ neurons ($F_{act}$ = 0.5) interacting with the astrocyte, reaches the threshold:
 
\begin{equation}
\label{eq:Nact}
\begin{aligned}
\frac{1}{N_{AE}}\sum_{(i,j)\in{N_{AE}}}\Theta([G]^{(i,j)} - [G]_{thr}) > F_{act}. 
\end{aligned}
\end{equation}
 
Thus, intracellular astrocytic calcium elevations occur in response to the increased concentration of the neurotransmitter released by excitatory neurons when a group of them fire coherently \cite{Perea2007, Navarrete2008, Navarrete2010}. In turn, gliotransmitters are released by activated astrocytes (Fig.~\ref{FigM1} E) modulating the strength of the synaptic connections in the corresponding neuronal group (Fig.~\ref{FigM1} C right):
 
\begin{equation}
\label{eq:astro_neuro}
\begin{aligned}
\overline{w_{synEE}} &=w_{synEE} \ \left(1+\nu_{Ca}\right),  w_{synEE} \in [10^{-4},w_{synEEmax}]\\
\nu_{Ca} &= \nu_{Ca}^*\Theta\left([Ca^{2+}]-[Ca^{2+}]_\text{thr} \right)\Theta\left(F - F_{astro}\right), \\
\end{aligned}
\end{equation} 
 
\noindent where $w_{synEE}$ is the excitatory synapse weight, trained according to the STDP rule (Fig.~\ref{FigM1} F), $\nu_{Ca}^*$ represents the strength of astrocytic modulation of synaptic transmission.
Astrocytic modulation of synaptic transmission is activated when the intracellular astrocytic $Ca^{2+}$ concentration exceeds the threshold $[Ca^{2+}]_\text{thr}$, and the proportion of active neurons, $F$, out of the total number of neurons $N_a$ associated with this astrocyte, exceeds the threshold $F_{astro}$. Such an organization of neuron-astrocyte interaction allows the astrocytes to integrate and coordinate a unique volume of synaptic activity. See Table~\ref{tab:SNAN_params} for model parameters.
 
\begin{table}[ht]
\centering
\caption{Spiking Neuron-Astrocyte Network Model Parameters}
\label{tab:SNAN_params}
\begin{tabular}{p{1.8cm}p{6.5cm}c}
\hline
\textbf{Component} & \textbf{Description} & \textbf{Value/Size} \\ \hline
\multicolumn{3}{c}{\textit{Neural Network Parameters}} \\ \hline
$W\times H$ & Excitatory neurons layer grid size & $81\times 81$ \\
$W_1\times H_1$ & Inhibitory neurons layer grid size & $40\times 40$ \\
$a$ & Time scale of the recovery variable & 0.1 \\
$b$ & Sensitivity of recovery variable to membrane potential fluctuations & 0.2 \\
$c$ & After-spike reset value of membrane potential & -65 mV \\
$d$ & After-spike reset value of recovery variable & 2 \\
$\eta$ & Baseline synaptic weight without astrocytic influence & 0.025 \\
$E_{\text{syn}}^e$ & Synaptic reversal potential (excitatory) & 0 mV \\
$E_{\text{syn}}^i$ & Synaptic reversal potential (inhibitory) & -90 mV \\
$k_{\text{syn}}$ & Slope of synaptic activation function & 0.2 mV \\ \hline
\multicolumn{3}{c}{\textit{Excitatory-Excitatory neuron Connections (EE)}} \\ \hline
$N_{\text{outEE}}$ & Output connections per neuron & 200 \\
$\lambda_{\text{EE}}$ & Exponential dist. rate for connection distance & 15 \\
$g_{synEE}$ & Weight change during learning & 0.007 \\
$\tau$ & Plasticity window & 20 ms \\
$w_{synEEmax}$ & Maximum synaptic weight & 0.05 \\ \hline
\multicolumn{3}{c}{\textit{Excitatory-Inhibitory neuron Connections (EI)}} \\ \hline
$N_{\text{outEI}}$ & Output connections per Excitatory neuron & 5 \\
$\lambda_{\text{EI}}$ & Exponential dist. rate for connection distance & 2 \\
$w_{synEI}$ & Fixed synaptic connection weight & 0.1 \\ \hline
\multicolumn{3}{c}{\textit{Inhibitory-Excitatory neuron Connections (IE)}} \\ \hline
$N_{\text{outIE}}$ & Output connections per Inhibitory neuron & 2000 \\
$\lambda_{\text{IE}}$ & Exponential dist. rate for connection distance & 80 \\
$g_{synIE}$ & Weight change during learning & 0.007 \\
$w_{synIEmax}$ & Maximum synaptic weight & 0.05 \\
$f^*$ & Firing rate threshold for IE plasticity & 0.3 \\ \hline
\multicolumn{3}{c}{\textit{Astrocytic Network Parameters}} \\ \hline
$M\times N$ & Astrocytic network grid size & $20\times 20$ \\
$c_0$ & Total Ca$^{2+}$ concentration & 2.0 $\mu$M \\
$c_1$ & ER-to-cytosol volume ratio & 0.185 \\
$v_1$ & Max Ca$^{2+}$ channel flux & 6 s$^{-1}$ \\
$v_2$ & Ca$^{2+}$ leak flux constant & 0.11 s$^{-1}$ \\
$v_3$ & Max Ca$^{2+}$ uptake rate & 2.2 $\mu$M s$^{-1}$ \\
$v_4$ & Max IP$_3$ production rate & 0.3 $\mu$M s$^{-1}$ \\
$v_6$ & Max activation-dependent Ca$^{2+}$ influx & 0.2 $\mu$M s$^{-1}$ \\
$k_1$ & Calcium extrusion rate constant & 0.5 s$^{-1}$ \\
$k_2$ & Half-saturation for Ca$^{2+}$ entry & 1 $\mu$M \\
$k_3$ & Activation constant for ATP-Ca$^{2+}$ pump & 0.1 $\mu$M \\
$d_1$ & IP$_3$ dissociation constant & 0.13 $\mu$M \\
$d_2$ & Ca$^{2+}$ inhibition constant & 1.049 $\mu$M \\
$d_3$ & Receptor dissociation constant & 943.4 nM \\
$d_5$ & Ca$^{2+}$ activation constant & 82 nM \\
$\alpha$ & IP$_3$ receptor binding parameter & 0.8 \\
$\tau_r^{-1}$ & IP$_3$ degradation rate & 0.14 s$^{-1}$ \\
$[IP_3]^*$ & Steady-state IP$_3$ concentration & 0.16 $\mu$M \\
$k_4$ & Ca$^{2+}$ stimulation constant & 1.1 $\mu$M \\
$d_{ca}$ & Ca$^{2+}$ diffusion rate & 0.05 s$^{-1}$ \\
$d_{ip3}$ & IP$_3$ diffusion rate & 0.05 s$^{-1}$ \\ \hline
\multicolumn{3}{c}{\textit{Neuron-Astrocyte Interaction}} \\ \hline
$N_{AE}$ & Neurons per astrocyte & 25 ($5\times 5$) \\
$\alpha_\text{glu}$ & Glutamate clearance constant & 50 s$^{-1}$ \\
$k_\text{glu}$ & Glutamate release efficacy & 600 $\mu$M s$^{-1}$ \\
$A_{glu}$ & IP$_3$ production rate via glutamate & 5 $\mu$M s$^{-1}$ \\
$t_\text{glu}$ & Glutamate stimulation duration & 60 ms \\
$G_\text{thr}$ & Glutamate threshold for IP$_3$ production & 0.2 \\
$F_\text{act}$ & Neuron sync fraction for Ca$^{2+}$ elevation & 0.75 \\
$F_\text{astro}$ & Neuron sync fraction for synaptic modulation & 0.5 \\
$\nu_{Ca}^*$ & Strength of astrocyte-induced modulation & 2 \\
$[Ca^{2+}]_\text{thr}$ & Ca$^{2+}$ threshold for modulation & 0.15 $\mu$M \\
$\tau_\text{astro}$ & Duration of synaptic modulation & 20 ms \\
\hline
\end{tabular}
\end{table}

\subsection{Convolutional Neural Network}\label{sec_Met_CNN}

\begin{figure}
\centering
\includegraphics[width=1\textwidth]{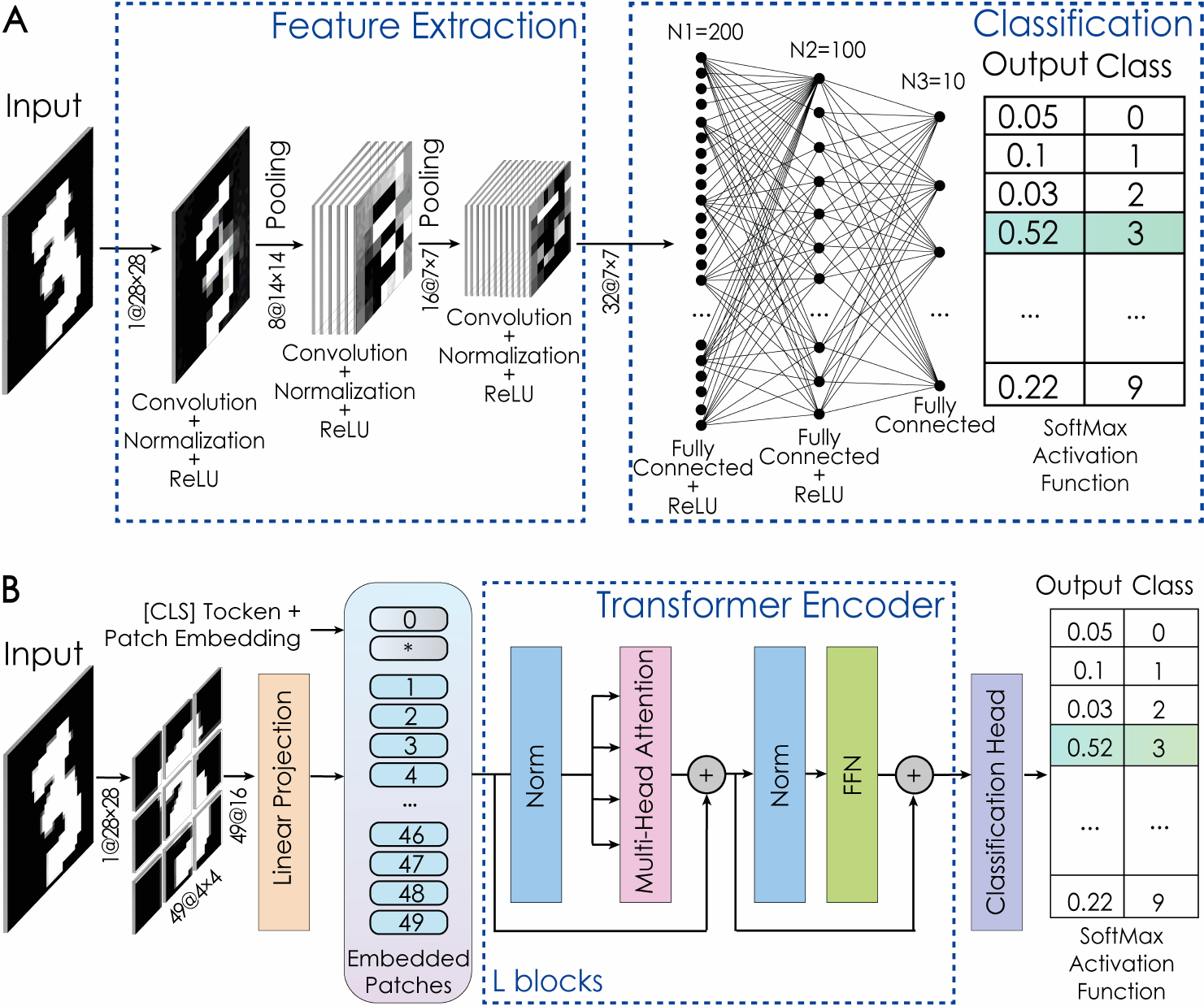}
\caption{(A) - Convolutional neural network (CNN) model topology. (B) - A Vision Transformer model topology.}
\label{FigM2}
\end{figure}

The CNN architecture is based on the LeNet-5 model \cite{LeCun1998} and is illustrated with Fig.~\ref{FigM2}A. The CNN accepts $28\times 28$ images normalized by dataset-specific mean subtraction. The architecture comprises three convolutional blocks: a $3\times 3$ convolution with 8 filters followed by batch normalization \cite{Ioffe2015}, ReLU activation \cite{Nair2010}, and $2\times 2$ max-pooling; an identical block with 16 filters; and a final $3\times 3$ convolutional layer with 32 filters, batch normalization, and ReLU. Spatial resolution reduces exponentially from $28\times 28$ to $7\times 7$ through two pooling operations. Resultant features are flattened and processed by fully-connected layers (200 $\to$ 100 $\to$ 10 units) with ReLU activations preceding the softmax output layer for class probability estimation. Training employed stochastic gradient descent with momentum ($\mu^{c}$=0.9) \cite{Sutskever2013} for 30 epochs using minibatches of 128 samples, an initial learning rate of 0.01 with exponential decay ($\gamma^{c}$=0.001), and random spatial translations ($\pm$5 pixels) for data augmentation \cite{Krizhevsky2012}.
 
The network comprises three convolutional blocks followed by fully-connected layers. Each convolutional block $k$ consists of:
\begin{align}
\mathbf{h}_k^{(1)} &= \text{conv2d}(\mathbf{h}_{k-1}, \mathbf{W}_k, \mathbf{b}_k) \quad \text{(with 'same' padding)} \\
\mathbf{h}_k^{(2)} &= \text{BatchNorm}(\mathbf{h}_k^{(1)}, \gamma_k, \beta_k) \\
\mathbf{h}_k^{(3)} &= \text{ReLU}(\mathbf{h}_k^{(2)}) \\
\mathbf{h}_k &= \text{maxpool}(\mathbf{h}_k^{(3)}, 2 \times 2, \text{stride}=2)
\end{align}
where $\mathbf{W}_k$ and $\mathbf{b}_k$  are learnable filters and biases, with $\gamma_k$ and $\beta_k$ being batch normalization parameters.
For the three convolutional blocks ($k=1,2,3$), the dimensions are:
\begin{align*}
\mathbf{W}_1 &\in \mathbb{R}^{8 \times 3 \times 3 \times 1}, \quad \mathbf{b}_1 \in \mathbb{R}^{8}, \quad \gamma_1,\beta_1 \in \mathbb{R}^{8};\\
\mathbf{W}_2 &\in \mathbb{R}^{16 \times 3 \times 3 \times 8}, \quad \mathbf{b}_2 \in \mathbb{R}^{16}, \quad \gamma_2,\beta_2 \in \mathbb{R}^{16};\\
\mathbf{W}_3 &\in \mathbb{R}^{32 \times 3 \times 3 \times 16}, \quad \mathbf{b}_3 \in \mathbb{R}^{32}, \quad \gamma_3,\beta_3 \in \mathbb{R}^{32}.
\end{align*}
 
\noindent\textbf{Layer Specifications:}
\begin{enumerate}
\item \textbf{Block 1:} 8 filters of size $3\times3$:
\begin{equation}
\mathbf{h}_0 = \mathbf{\hat{I}}, \quad \mathbf{h}_1^{(1)} \in \mathbb{R}^{28\times28\times8}
\end{equation}
After max-pooling: $\mathbf{h}_1 \in \mathbb{R}^{14\times14\times8}$
 
\item \textbf{Block 2:} 16 filters of size $3\times3$:
\begin{equation}
\mathbf{h}_2^{(1)} \in \mathbb{R}^{14\times14\times16}, \quad \mathbf{h}_2 \in \mathbb{R}^{7\times7\times16}
\end{equation}
 
\item \textbf{Block 3:} 32 filters of size $3\times3$ (without pooling):
\begin{equation}
\mathbf{h}_3^{(1)} \in \mathbb{R}^{7\times7\times32}, \quad \mathbf{h}_3 \in \mathbb{R}^{7\times7\times32}
\end{equation}
\end{enumerate}
 
The output is flattened and processed through fully-connected layers:
\begin{align}
\mathbf{v^{c}} &= \text{flatten}(\mathbf{h}_3) \in \mathbb{R}^{1568} \\
\mathbf{v}_1^{c} &= \text{ReLU}(\mathbf{W}_4 \mathbf{v^{c}} + \mathbf{b}_4), \quad \mathbf{W}_4 \in \mathbb{R}^{200 \times 1568} \\
\mathbf{v}_2^{c} &= \text{ReLU}(\mathbf{W}_5 \mathbf{v}_1^{c} + \mathbf{b}_5), \quad \mathbf{W}_5 \in \mathbb{R}^{100 \times 200} \\
\mathbf{y^{c}} &= \mathbf{W}_6 \mathbf{v}_2^{c} + \mathbf{b}_6, \quad \mathbf{W}_6 \in \mathbb{R}^{10 \times 100} \\
\mathbf{p} &= \text{softmax}(\mathbf{y^{c}})
\end{align}

\subsubsection{Model Initialization}
The CNN was implemented using the MATLAB Deep Learning Toolbox. All convolutional layer weights were initialized using the Glorot (Xavier) uniform distribution \cite{Glorot2010}, which draws values from $\mathcal{U}(-\sqrt{6/(n_{in}+n_{out})},\, \sqrt{6/(n_{in}+n_{out})})$, where $n_{in}$ and $n_{out}$ denote the number of input and output units, respectively. Biases were initialized to zero. Batch normalization scale parameters ($\gamma_k$) were initialized to 1 and offset parameters ($\beta_k$) to 0. Fully-connected layer weights were also initialized via the Glorot scheme with zero biases.
 
Input calibration is performed by subtracting the dataset mean:
\begin{equation}
\mathbf{\hat{I}} = \mathbf{I} -  \bar{\mathbf{I}}, \quad \bar{\mathbf{I}} = \frac{1}{N_{\text{train}}} \sum_{i=1}^{N_{\text{train}}} \mathbf{I}_i
\end{equation}
 
\begin{table}[ht]
\centering
\caption{Convolutional Neural Network Model Parameters}
\label{tab:cnn_params}
\begin{tabular}{p{2.2cm}p{6.2cm}c}
\hline
\textbf{Component} & \textbf{Description} & \textbf{Value/Size} \\ \hline
\multicolumn{3}{c}{\textit{Architecture Parameters}} \\ \hline
$\mathbf{I}$ & Input image size (MNIST) & $28 \times 28 \times 1$ \\
Conv1 & Filters, kernel, padding & 8, $3\times3$, 'same' \\
Pool1 & Pooling type and size & MaxPool, $2\times2$ \\
Conv2 & Filters, kernel, padding & 16, $3\times3$, 'same' \\
Pool2 & Pooling type and size & MaxPool, $2\times2$ \\
Conv3 & Filters, kernel, padding & 32, $3\times3$, 'same' \\
FC1 & Hidden units & 200 \\
FC2 & Hidden units & 100 \\
Output & Classes & 10 \\
Activation & Non-linearity & ReLU \\ \hline
\multicolumn{3}{c}{\textit{Training Hyperparameters}} \\ \hline
Batch size & Samples per batch & 128 \\
Optimizer & Optimization algorithm & SGDM \\
Momentum ($\mu^{c}$) & Momentum coefficient & 0.9 \\
Initial LR ($\eta_0^{c}$) & Learning rate & 0.01 \\
LR decay ($\gamma^{c}$) & Exponential decay & 0.001 \\
Epochs & Training iterations & 30 \\
Augmentation & Spatial translations & $\pm5$ pixels \\
\hline
\end{tabular}
\end{table}

\subsection{Vision Transformer Model}\label{sec_ViT}
The Vision Transformer (ViT) model employed in this work follows the seminal architecture introduced by Dosovitskiy et al. \cite{Dosovitskiy2021}, adapted to our experimental setting. The architecture of the ViT model is illustrated in Fig.~\ref{FigM2}B. It processes $28\times28$ images partitioned into 49 non-overlapping $4\times4$ patches. Each patch is embedded into a 128-dimensional space ($D^{t}=128$) and augmented with a learnable [CLS] token \cite{Devlin2019} and positional embeddings \cite{Vaswani2017}. The sequence is processed by $L^{t}=8$ transformer layers, each containing Multi-Head Self-Attention (MSA) with $H^{t}=8$ heads (each with key/query/value dimension $d_k = 64$) and a ReLU-activated Feed-Forward Network (FFN) with hidden dimension 256. The final [CLS] token state is used for classification via a linear projection to 10 classes. Training employed the Adam optimizer \cite{Kingma2014} with a learning rate of $5 \times 10^{-4}$ for 50 epochs using minibatches of 128 samples. See Table~\ref{tab:vit_params} for model parameters.
 
\subsubsection{Model Initialization}
The ViT was implemented in PyTorch. The patch projection matrix $\mathbf{E^{t}}$ and all linear layers in the attention and feed-forward modules were initialized using PyTorch's default Kaiming uniform initialization \cite{He2015}. The learnable [CLS] token $\mathbf{z}_{\text{cls}} \in \mathbb{R}^{128}$ and positional embeddings $\mathbf{E}_{\text{pos}} \in \mathbb{R}^{50 \times 128}$ were initialized by sampling from a standard normal distribution $\mathcal{N}(0, 1)$. All bias terms were initialized to zero by default, and layer normalization parameters were initialized with unit scale and zero offset.
 
The input image ($\mathbf{I} \in \mathbb{R}^{28 \times 28 \times 1}$) is partitioned into $N^{t}$ = 49 non-overlapping $4\times 4$ patches ($\mathbf{I_p} \in \mathbb{R}^{16}$). Each patch is linearly projected to a $D^{t}$-dimensional feature space ($D^{t}=128$) via a trainable matrix $\mathbf{E^{t}} \in \mathbb{R}^{16 \times 128}$, yielding patch embeddings $\mathbf{z}_0 = \left[ \mathbf{I}_p^{(1)} \mathbf{E^{t}}, \mathbf{I}_p^{(2)} \mathbf{E^{t}}, \dots, \mathbf{I}_p^{(N^{t})} \mathbf{E^{t}} \right]$.
 
A learnable \texttt{[CLS]} token $\mathbf{z}_{\text{cls}} \in \mathbb{R}^{D^{t}}$ is prepended to the patch sequence to aggregate global image representations. Positional embeddings $\mathbf{E}_{\text{pos}} \in \mathbb{R}^{(N^{t}+1) \times D^{t}}$ are added to retain spatial information:
 
\begin{equation}
\label{eq:T_1}
\begin{aligned}
\mathbf{z}_0 = \left[ \mathbf{z}_{\text{cls}}; \mathbf{I}_p^{(1)} \mathbf{E^{t}}; \dots; \mathbf{I}_p^{(N^{t})} \mathbf{E^{t}} \right] + \mathbf{E}_{\text{pos}}.\\
\end{aligned}
\end{equation} 
 
\subsubsection{Transformer Encoder}
The embedded sequence $\mathbf{z}_0$ is processed by $L^{t} = 8$ identical transformer layers. Each layer consists of two sub-layers (Multi-Head Self-Attention (MSA) and Feed-Forward Network (FFN)):
 
\begin{equation}
\label{eq:T_2}
\begin{aligned}
\mathbf{z}'_l &= \operatorname{MSA}(\operatorname{LayerNorm}(\mathbf{z}_{l-1})) + \mathbf{z}_{l-1} \\
\mathbf{z}_l &= \operatorname{FFN}(\operatorname{LayerNorm}(\mathbf{z}'_l)) + \mathbf{z}'_l
\end{aligned}
\end{equation}
\noindent where $l = 1, \dots, L^{t}$, with LayerNorm applied before each operation (PreNorm configuration).
 
The MSA sub-layer employs $H^{t} = 8$ parallel attention heads. Each head computes:
 
\begin{equation}
\label{eq:T_3}
\begin{aligned}
\operatorname{head}_h = \operatorname{Attention}\left( 
    \mathbf{Q^{t}}_h = \mathbf{z}\mathbf{W}_h^Q, 
    \mathbf{K^{t}}_h = \mathbf{z}\mathbf{W}_h^K, 
    \mathbf{V^{t}}_h = \mathbf{z}\mathbf{W}_h^V 
\right)
\end{aligned}
\end{equation}
 
\noindent where $\mathbf{W}_h^Q, \mathbf{W}_h^K, \mathbf{W}_h^V \in \mathbb{R}^{128 \times 64}$, and:
 
\begin{equation}
\label{eq:T_4}
\begin{aligned}
\operatorname{Attention}(\mathbf{Q^{t}},\mathbf{K^{t}},\mathbf{V^{t}}) = \operatorname{softmax}\left( \frac{\mathbf{Q^{t}}\mathbf{K^{t}}^\top}{\sqrt{d_k}} \right)\mathbf{V^{t}}
\end{aligned}
\end{equation} 
 
with $d_k = 64$. Outputs are concatenated and projected:
 
\begin{equation}
\label{eq:T_5}
\begin{aligned}
\operatorname{MSA}(\mathbf{z}) = \operatorname{Concat}(\operatorname{head}_1, \dots, \operatorname{head}_8) \mathbf{W}^O, \quad \mathbf{W}^O \in \mathbb{R}^{512 \times 128}
\end{aligned}
\end{equation} 
 
The FFN sub-layer operates identically on each position:
\begin{equation}
\label{eq:T_6}
\begin{aligned}
\operatorname{FFN}(\mathbf{z}) = \operatorname{ReLU}\left( \mathbf{z} \mathbf{W^{t}}_1 + \mathbf{b^{t}}_1 \right) \mathbf{W^{t}}_2 + \mathbf{b^{t}}_2
\end{aligned}
\end{equation}
 
\noindent where $\mathbf{W^{t}}_1 \in \mathbb{R}^{128 \times 256}$, $\mathbf{W^{t}}_2 \in \mathbb{R}^{256 \times 128}$. Unlike ViT from \cite{Dosovitskiy2021} which uses GELU, we employ ReLU activation for computational efficiency and training stability. On small and simple datasets such as MNIST and Fashion-MNIST the accuracy difference between ReLU and GELU is minimal, and the dying ReLU problem is mitigated by proper initialization \cite{He2015}.
 
\subsubsection{Classification Head}
The classification head processes the final state of the \texttt{[CLS]} token ($\mathbf{z}_L^0$), which serves as a global image representation. First, layer normalization is applied to stabilize feature scales:
 
\begin{equation}
\label{eq:T_7}
\begin{aligned}
\mathbf{h^{t}} = \operatorname{LayerNorm}(\mathbf{z}_L^0), \quad \mathbf{h^{t}} \in \mathbb{R}^{128}
\end{aligned}
\end{equation}
 
This normalized representation is then projected to class logits via a linear transformation:
\begin{equation}
\label{eq:T_8}
\begin{aligned}
\mathbf{y^{t}} = \mathbf{W^{t}}_c \mathbf{h^{t}} + \mathbf{b^{t}}_c, \quad \mathbf{W^{t}}_c \in \mathbb{R}^{10 \times 128}, \: \mathbf{b^{t}}_c \in \mathbb{R}^{10}
\end{aligned}
\end{equation}
 
\noindent where $\mathbf{y^{t}}$ contains unnormalized log probabilities for each of the 10 classes. Final class probabilities are obtained through softmax normalization:
\begin{equation}
\label{eq:T_9}
\begin{aligned}
\mathbf{p} = \operatorname{softmax}(\mathbf{y^{t}})
\end{aligned}
\end{equation}
 
\begin{table}[ht]
\centering
\caption{Vision Transformer model parameters}
\label{tab:vit_params}
\begin{tabular}{p{2.2cm}p{6.2cm}c}
\hline
\textbf{Component} & \textbf{Description} & \textbf{Value/Size} \\ \hline
\multicolumn{3}{c}{\textit{Architecture Parameters}} \\ \hline
$I$ & Input image size (MNIST) & $28 \times 28 \times 1$ \\
$P$ & Patch size & $4 \times 4$ \\
$N^{t}$ & Number of patches & $49$ \\
$D^{t}$ & Patch embedding dimension & $128$ \\
$\mathbf{E^{t}}$ & Patch projection matrix & $\mathbb{R}^{16 \times 128}$ \\
$L^{t}$ & Number of transformer layers & $8$ \\
$H^{t}$ & Number of attention heads & $8$ \\
$d_k$ & Key/query/value dimension per head & $64$ \\
$\mathbf{W}^O$ & MSA output projection matrix & $\mathbb{R}^{512 \times 128}$ \\
FFN hidden & FFN hidden layer size & $256$ \\
FFN activation & FFN activation function & ReLU \\
$\mathbf{W^{t}}_c$ & Classifier weights & $\mathbb{R}^{10 \times 128}$ \\ \hline
\multicolumn{3}{c}{\textit{Training Hyperparameters}} \\ \hline
Batch size & Samples per training batch & 128 \\
Optimizer & Optimization algorithm & Adam \\
Learning rate & Optimizer step size & $5 \times 10^{-4}$ \\
Epochs & Training iterations & 50 \\
\hline
\end{tabular}
\end{table}

\subsection{Experimental Setup}\label{train_test}
\begin{figure}
\centering
\includegraphics[width=1\textwidth]{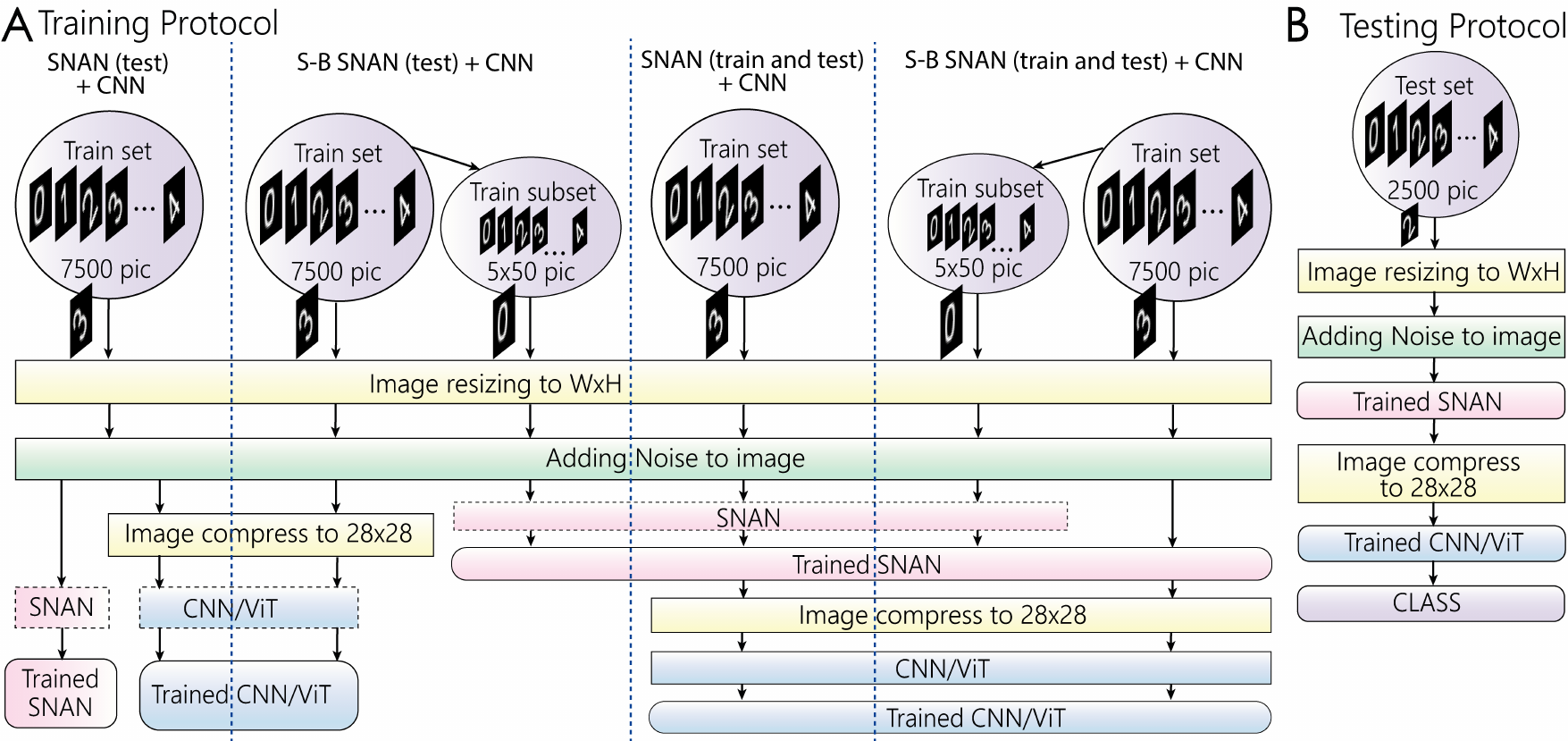}
\caption{(A) - SNAN-CNN training protocol.(B) - SNAN-CNN testing protocol.}
\label{FigM3}
\end{figure}

We evaluated four SNAN integration strategies coupled with two distinct classifier architectures---a convolutional neural network (CNN) and a Vision Transformer (ViT)---during training and testing (Fig.~\ref{FigM3}). For brevity, the protocols are described below for the CNN case; identical configurations were used with the ViT classifier.

\textit{1. SNAN (test) + CNN}

Both models were trained separately on the MNIST training subset (7500 images). The SNAN was trained with a single presentation per image, using Hebbian STDP for synaptic updates and forming stimulus-specific astrocytic calcium patterns. Its final synaptic weights and calcium concentrations were fixed for inference. The CNN was trained separately for 30 epochs via stochastic gradient descent with momentum ($\mu^{c}=0.9$; Section~\ref{sec_Met_CNN}). During testing, all 2500 test images were sequentially processed: first by the SNAN to produce mean excitatory firing rates, which were min-max normalized to $[0,1]$ and resampled to $28\times28$ resolution, then classified by the CNN.

\textit{2. SNAN (train and test) + CNN}

The SNAN was first trained and fixed as in Variant 1. The CNN was then trained on the entire MNIST training set preprocessed by this fixed SNAN (i.e., on normalized and resampled firing rate maps). Inference was identical to Variant 1.

\textit{3. S-B SNAN (test) + CNN}

The situation-based (S-B) SNAN \cite{Gordleeva2023} was trained on a situational subset of 250 images (5 classes, 50 samples/class) to form situation-specific calcium patterns. Its parameters were fixed post-training. The CNN was trained on the full MNIST training set (7500 images). During inference, only test samples from the 5 target classes (1250 images) were processed through the fixed S-B SNAN (generating normalized firing rate maps) and then classified by the CNN.

\textit{4. S-B SNAN (train and test) + CNN}

The S-B SNAN was trained and fixed on the situational subset (250 images) as in Variant 3. The CNN was then trained on the entire MNIST training set (7500 images), but each image was first preprocessed by the fixed S-B SNAN. Inference was performed identically to Variant 3, processing only the 1250 situation-specific test images.

We also evaluated two baseline models: a standalone CNN (Methods~\ref{sec_Met_CNN}) and a Vision Transformer (Methods~\ref{sec_ViT}). The CNN was trained for 30 epochs on raw MNIST training images (7500 samples) using stochastic gradient descent with momentum ($\mu^{c}=0.9$) and evaluated on the unprocessed test set (2500 samples). The ViT was trained end-to-end for 50 epochs on the same unprocessed training set using Adam optimization \cite{Kingma2014} with a learning rate of $5 \times 10^{-4}$ and evaluated on all test samples.

\subsection{Image Binarization}\label{sec:binarization}
Before being fed into the SNAN, all input images from the MNIST and Fashion-MNIST datasets were binarized using a fixed global threshold. Each grayscale image was resized to SNAN Input size ($W\times H$ pixels), and pixels with intensity below 50 were set to 0 (background), while all other pixels were set to 1 
(object). This simple operation transforms the original 8‑bit input into a binary matrix. Binarization is known to reduce the amount of information available to a classifier, which generally leads to a drop in classification accuracy compared to using full-precision data~\cite{Kayumov2021}. Despite the loss of grayscale information, the binary representation preserves the essential shape features required for classification and matches the SNAN's input format.
 
\subsection{Adding Noise to Images}\label{sec_Met_noise}
 
To evaluate denoising performance under diverse corruption scenarios, two distinct noise types were applied to test images after resizing them to match the SNAN excitatory layer dimensions:
 
\subsubsection{Random ”Salt and pepper” Impulse Noise}\label{sec_rand_noise}
 
We applied impulse noise by randomly inverting pixel intensities. For each image, a fraction of pixels ($Noise$ $level$) were selected uniformly at random. Representative examples of corrupted images across noise levels are shown in Fig.~\ref{Fig2}C.
 
\subsubsection{Partial Occlusion}\label{sec_occl}

\noindent\textbf{Test-time occlusion.} To evaluate robustness against realistic visual obstructions, we occluded test images by overlaying randomly selected digit patterns from the \emph{test subset}. Using test images as occlusion masks guarantees that the model has never seen these patterns during training, preventing data leakage and providing an unbiased estimate of occlusion robustness. Figure~\ref{Fig2}D illustrates representative examples for four occlusion sizes (20\%, 40\%, 60\%, 80\%).

The procedure was:
\begin{enumerate}
    \item An occlusion image $I_d$ was randomly selected from the test subset
    \item $I_d$ was scaled to $Occlusion$ $size,\%$ of the target image size
    \item The scaled $I_d$ was positioned at one of four corners (randomly chosen)
    \item Intensity normalization preserved original dynamic range:
    \begin{equation}
    I_{\text{combined}} = \min\left( I_{\text{original}} + I_d,  1 \right)
    \end{equation}
\end{enumerate}
 
\noindent\textbf{Training-time Augmentation.} To enhance model robustness to partial occlusions, we augmented the training set with random scaling and positioning variations. The augmentation parameters were varied across experiments: the minimum scale factor $size_{\min}$ was set to $10\%$, $40\%$, or $65\%$ of the original image size (see Results,  Section~\ref{subsec2_4_2}). For each original $28\times28$ image, the procedure was:
\begin{enumerate}
    \item Randomly select a scale factor $s \sim \mathcal{U}[size_{\min}, 100\%]$ (i.e., between $size_{\min}\%$ and $100\%$ of the original size).
    \item Rescale the image to $X = \lfloor 28 \cdot s/100 \rfloor$, $Y = \lfloor 28 \cdot s/100 \rfloor$ pixels using nearest-neighbor interpolation.
    \item Place the resized image on a $28\times28$ zero canvas at a random position $(x,y)$, where $x \sim \mathcal{U}[0, 28-X]$, $y \sim \mathcal{U}[0, 28-Y]$.
\end{enumerate}
This transformation simulates variations in object scale and position, forcing the model to rely on local discriminative features and improving its ability to recognise partially occluded digits. Representative samples are shown in Fig.~\ref{Fig2}E.
 
\subsection{CNN with Classical Image Preprocessing}\label{sec_filters}
 
Three denoising pipelines combining conventional image filters with the standard CNN classifier:
 
\subsubsection{Median Filter \cite{Huang1981} + CNN}
 
Test images were processed with a two-dimensional median filter using a $5\times5$ pixel neighborhood. Each output pixel value is computed as:
\begin{equation}
I_{\text{out}}(x,y) = \underset{(i,j) \in N_{5\times5}(x,y)}{\text{median}} I_{\text{in}}(i,j)
\end{equation}
 
\noindent where $N_{5\times5}(x,y)$ denotes the $5\times5$ window centered at $(x,y)$.
    
\subsubsection{Mean Filter \cite{Pratt2006} + CNN}
 
Spatial averaging was applied using a $6\times6$ uniform kernel with symmetric boundary handling. The convolution operation is defined as:
\begin{equation}
I_{\text{out}}(x,y) = \frac{1}{36} \sum_{i=-3}^{2} \sum_{j=-3}^{2} I_{\text{in}}(x+i,y+j)
\end{equation}

\subsubsection{Bilateral Filter \cite{Tomasi1998} + CNN}
 
Test images underwent edge-preserving denoising via a bilateral filter with spatial parameter $\sigma_s=2$ and range parameter $\sigma_r=50$. The filtered intensity $I_{\text{out}}(\mathbf{x})$ at pixel position $\mathbf{x}$ is computed as:
 
\begin{equation}
I_{\text{out}}(\mathbf{x}) = \frac{1}{W^f(\mathbf{x})} \sum_{\mathbf{x}_i \in \Omega} I_{in}(\mathbf{x}_i) \cdot w_s(\|\mathbf{x}_i - \mathbf{x}\|) \cdot w_r(|I_{in}(\mathbf{x}_i) - I_{in}(\mathbf{x})|)
\end{equation}
 
\noindent where $w_s(\|\mathbf{x}_i - \mathbf{x}\|) = \exp\left(-\frac{\|\mathbf{x}_i - \mathbf{x}\|^2}{2\sigma_s^2}\right)$ is the spatial weighting kernel; $w_r(|I_{in}(\mathbf{x}_i) - I_{in}(\mathbf{x})|) = \exp\left(-\frac{|I_{in}(\mathbf{x}_i) - I_{in}(\mathbf{x})|^2}{2\sigma_r^2}\right)$ is the range weighting kernel; $W^f(\mathbf{x}) = \sum_{\mathbf{x}_i \in \Omega} w_s(\|\mathbf{x}_i - \mathbf{x}\|) \cdot w_r(|I_{in}(\mathbf{x}_i) - I_{in}(\mathbf{x})|)$ is the normalization factor; $\Omega$ represents the spatial neighborhood around $\mathbf{x}$.
 
In all cases, filtered test images were resampled to $28\times28$ resolution via bicubic interpolation before CNN processing. Crucially, no filtering was applied during CNN training---the classifier was trained exclusively on unfiltered images from the MNIST training set. The underlying CNN architecture and weights remained identical to the standalone CNN baseline (Model 5).

\subsection{Cluster Metrics of Learned Representations}\label{sec_cluster_metrics}
Identical metrics were computed for both architectures to quantitatively evaluate clustering properties of learned representations. For the CNN, 100-dimensional embeddings from the $FC2$ layer were used. For the ViT, 128-dimensional embeddings from the final [CLS] token state (after the LayerNorm in the classification head) were used.
 
\subsubsection{Cluster Size} 
For each class $c$, the  cluster centroid $\mathbf{\mu}_c$ was computed as the arithmetic mean of all embeddings belonging to that class. The cluster size $C_s$ (a measure of its compactness) was then quantified as the mean Euclidean distance from each embedded point to its class centroid:
 
\begin{equation}
C_s^{(c)} = \frac{1}{N_c} \sum_{i=1}^{N_c} \|\mathbf{x}_i^{(c)} - \mathbf{\mu}_c \|
\end{equation}
where $N_c$ denotes the sample count for class $c$, $\mathbf{x}_i^{(c)}$  denotes the $i$-th embedding vector belonging to class $c$. Lower values of $C_s^{(c)}$ indicate tighter, more compact clusters.
 
\subsubsection{Inter-cluster Distance} 
Pairwise Euclidean distances between clusters $C_d$ were computed as:
\begin{equation}
C_d^{(j,k)} = \|\mathbf{\mu}_j - \mathbf{\mu}_k \| \quad \forall  j \neq k
\end{equation}
with higher values signifying better class separation.

\subsubsection{Cluster Metrics Normalization} 
To ensure scale-invariant comparisons, all distances were normalized by the total variability of the feature space:
\begin{enumerate}
    \item \textit{Centering:} The global mean $\bar{\mathbf{x}} = \frac{1}{N_{\text{test}}} \sum_{i=1}^{N_{\text{test}}} \mathbf{x}_i$ was computed, and embeddings were centered as:
    \begin{equation}
    \mathbf{x}_i^{\text{cent}} = \mathbf{x}_i - \bar{\mathbf{x}},
    \end{equation}
where $N_{\text{test}}$ is total number of samples in test set.
   
    \item \textit{Total Variance Scaling:} To quantify the total variability of the feature space, the covariance matrix was estimated from the centered embeddings:
    \begin{equation}
\mathbf{\Sigma} = \frac{1}{N_{\text{test}}-1} \sum_{i=1}^{N_{\text{test}}} \mathbf{x}_i^{\text{cent}} (\mathbf{x}_i^{\text{cent}})^\top
\end{equation}
 
Eigendecomposition of the covariance matrix was performed: $\mathbf{\Sigma} \mathbf{v}_j = \lambda_j \mathbf{v}_j$. The scaling factor $sf$ was defined as the square root of the total variance, which is given by the sum of all eigenvalues:
\begin{equation} \label{eq:sf}
sf = \sqrt{\sum_{j=1}^d \lambda_j}
\end{equation}
where $d$ is the dimensionality of the feature space.
 
    \item \textit{Normalization:} 
The cluster sizes and inter-cluster distances were averaged across classes and scaled by the scaling factor $sf$:
    \begin{align}
   \bar C_s &= \frac{1}{N_c} \sum_{c=1}^{N_c} \frac{C_s^{(c)}}{sf} \\
\bar C_d &= \frac{1}{N_c(N_c - 1)} \sum_{j=1}^{N_c} \sum_{\substack{k=1 \\ k \neq j}}^{N_c} \frac{C_d^{(j,k)}}{sf}
    \end{align}
\end{enumerate}
where $N_c$ is the number of classes.

\subsubsection{Cluster Shift}
To quantify the stability of the learned feature‑space geometry under increasing corruption, we introduce the cluster shift metric. Let $\boldsymbol{\mu}_c^{(0)}$ denote the centroid of class $c$ computed from clean test images (noise level 0), and $\boldsymbol{\mu}_c^{(\nu)}$ the centroid of the same class at noise level $\nu$. The displacement of class $c$ at noise level $\nu$ is defined as the Euclidean distance between these two centroids:
\begin{equation}
\Delta_c^{(\nu)} = \|\boldsymbol{\mu}_c^{(\nu)} - \boldsymbol{\mu}_c^{(0)}\|.
\end{equation}
To obtain a single, scale‑invariant measure per noise level, the displacements are first averaged over all classes and then normalised by the same scaling factor $sf$ introduced in Eq.~(\ref{eq:sf}):
\begin{equation}
\text{Shift}^{(\nu)} = \frac{1}{N_c} \sum_{c=1}^{N_c} \frac{\Delta_c^{(\nu)}}{sf},
\end{equation}
where $N_c$ is the number of classes. This value is computed for each random train–test split, and we report the mean and standard deviation across splits. Lower values of $\text{Shift}^{(\nu)}$ indicate that the class centroids remain close to their clean‑data positions, i.e., the feature‑space organisation is robust to noise.

\subsection{Fashion-MNIST and Omniglot Experimental Configurations}
\label{sec:fashion_omniglot}

\textbf{Fashion-MNIST.} For Fashion-MNIST we adopted the same training and testing splits as for MNIST: 750 images per class for training and 250 for testing. All images were binarised before processing. In occlusion experiments, each source image was first rescaled to 80\% of its original size and positioned on a blank background. Only then were the standard scale‑and‑shift augmentation, noise addition, and partial occlusion applied. This procedure ensures that the occluding object does not cover the entire target region and that the experimental conditions match those used for MNIST.

\textbf{Omniglot.} The Omniglot dataset consists of $105\times105$ pixel binary images of handwritten characters drawn from 50 different alphabets. For our evaluation we randomly selected 10 classes, each containing exactly 20 examples. To test few‑shot learning performance, we split the 20 examples per class into training and test sets in three configurations: 5, 10, or 15 training samples per class, with the remaining 15, 10, or 5 samples, respectively, used for testing.

To handle the larger input dimensions, the CNN architecture was modified to accept $105\times105$ pixel inputs. The vision transformer received images of the same size, with the patch size increased from $4\times4$ to $5\times5$. The SNAN dimensions were scaled accordingly: the excitatory neuron layer was set to $105\times105$. The astrocytic network, which for the original MNIST setup comprised $20\times20$ cells, was correspondingly expanded to $26\times26$ for the Omniglot experiments.

\begin{appendices}

\section{Additional figures}\label{secA}

\begin{figure}
\centering
\includegraphics[width=1\textwidth]{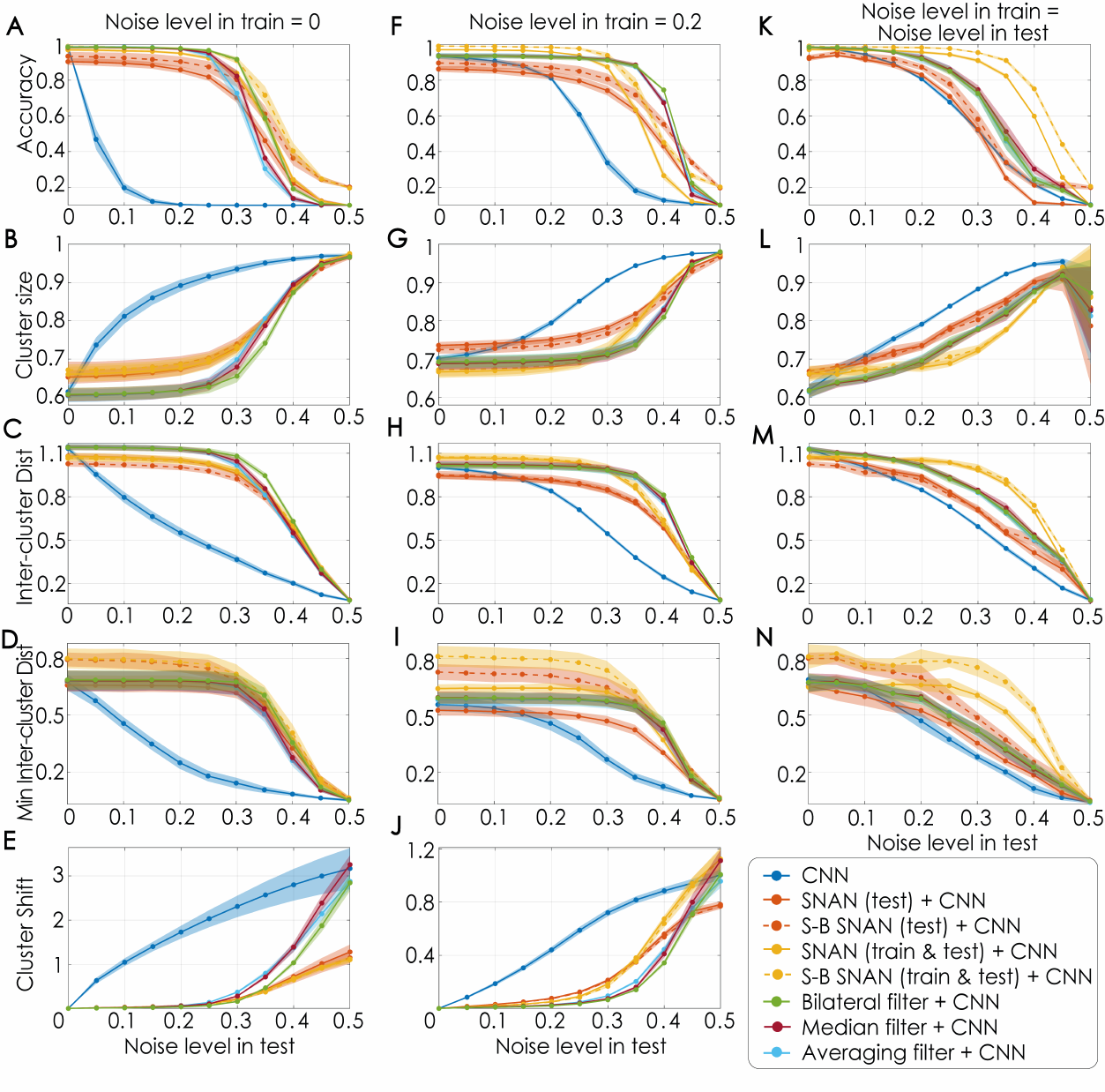}
\caption{Evolution of classification accuracy and feature space properties under increasing impulse noise during testing on the MNIST dataset for CNN-based models trained under different noise conditions. (A-E) Models trained on clean images. (F-J) Models trained with a fixed noise level ($Noise{\text{ }level} = 0.2$). (K-N) Models trained with noise level matched to the test condition. (A, F, K) Classification accuracy. (B, G, L) Normalized cluster size of learned representations, $\bar{C_s}$. (C, H, M) Pairwise inter-cluster distance, $\bar{C_d}$. (D, I, N) Nearest-class inter-cluster distance. (E, J) Cluster shift. All results are averaged over 10 runs; the solid lines represent the mean and the shaded regions correspond to $\pm$ SD.}
\label{FigS1}
\end{figure}

\begin{figure}
\centering
\includegraphics[width=1\textwidth]{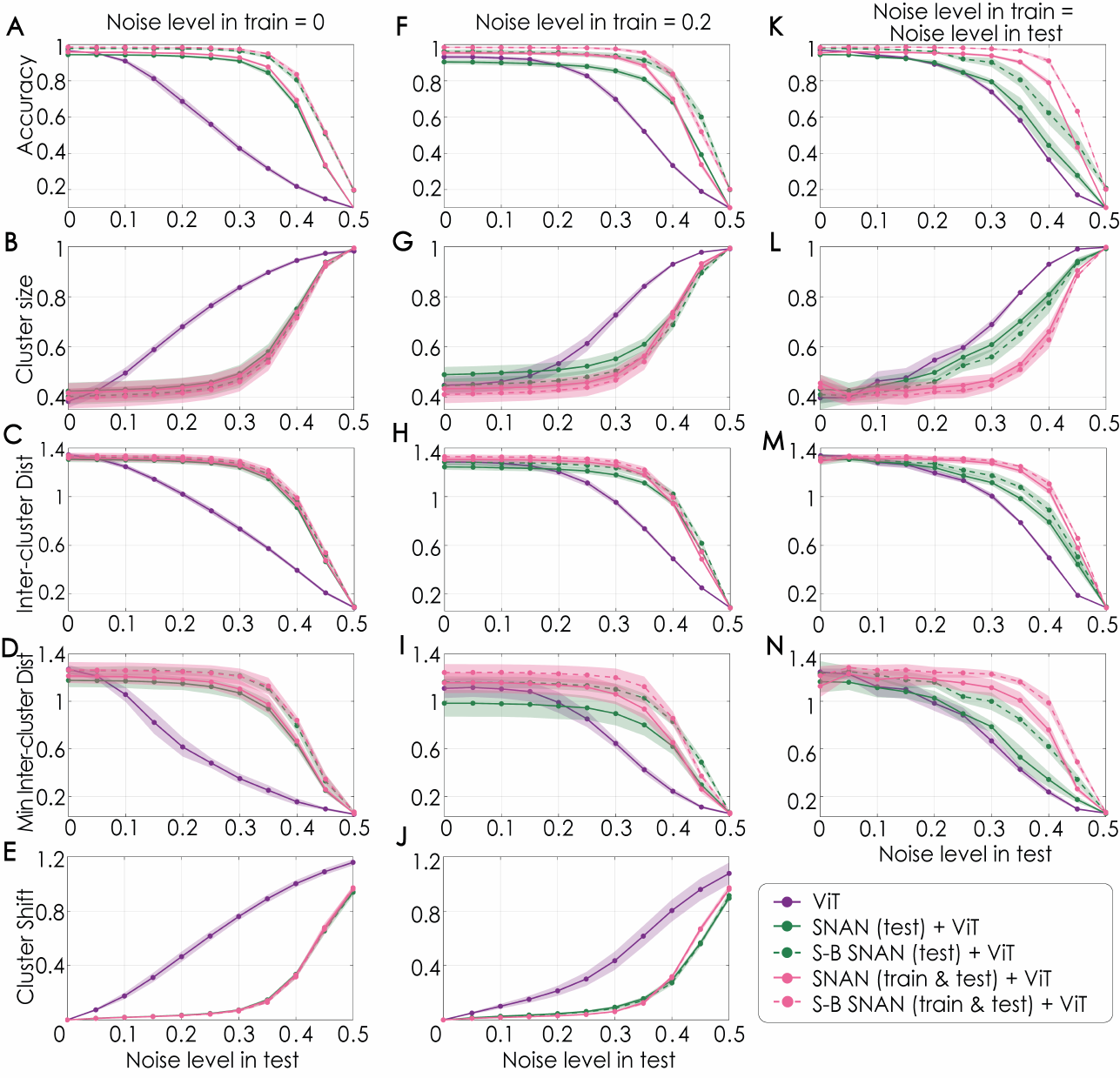}
\caption{Evolution of classification accuracy and feature space properties under increasing impulse noise during testing on the MNIST dataset for ViT-based models trained under different noise conditions. (A-E) Models trained on clean images. (F-J) Models trained with a fixed noise level ($Noise{\text{ }level} = 0.2$). (K-N) Models trained with noise level matched to the test condition. (A, F, K) Classification accuracy. (B, G, L) Normalized cluster size of learned representations, $\bar{C_s}$. (C, H, M) Pairwise inter-cluster distance, $\bar{C_d}$. (D, I, N) Nearest-class inter-cluster distance. (E, J) Cluster shift. All results are averaged over 10 runs; the solid lines represent the mean and the shaded regions correspond to $\pm$ SD.}
\label{FigS2}
\end{figure}

\begin{figure}
\centering
\includegraphics[width=1\textwidth]{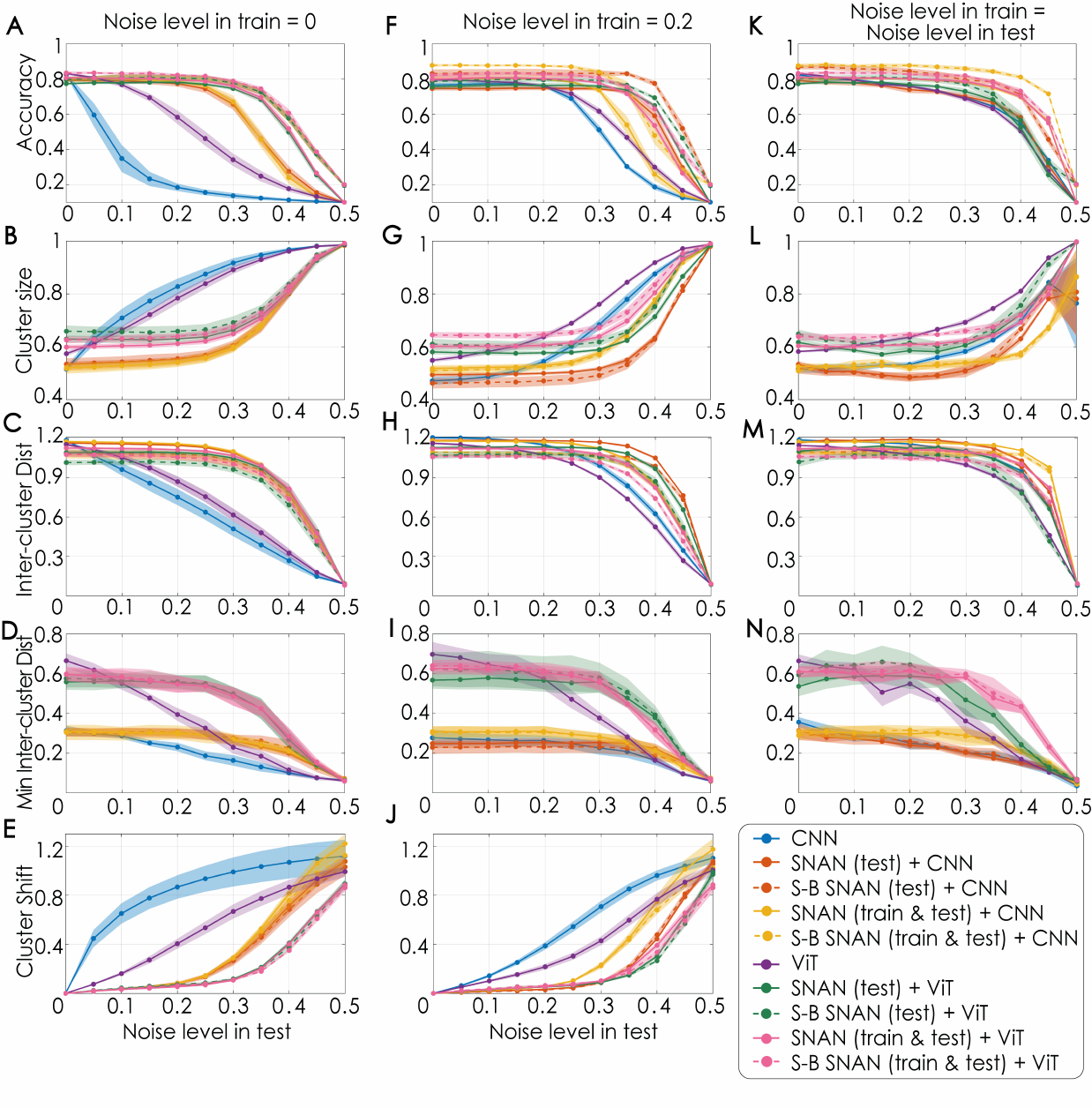}
\caption{Evolution of classification accuracy and feature space properties under increasing impulse noise during testing on the Fashion-MNIST dataset for models trained under different noise conditions. (A-E) Models trained on clean images. (F-J) Models trained with a fixed noise level ($Noise{\text{ }level} = 0.2$). (K-N) Models trained with noise level matched to the test condition. (A, F, K) Classification accuracy. (B, G, L) Normalized cluster size of learned representations, $\bar{C_s}$. (C, H, M) Pairwise inter-cluster distance, $\bar{C_d}$. (D, I, N) Nearest-class inter-cluster distance. (E, J) Cluster shift. All results are averaged over 10 runs; the solid lines represent the mean and the shaded regions correspond to $\pm$ SD.}
\label{FigS5}
\end{figure}

\begin{figure}
\centering
\includegraphics[width=1\textwidth]{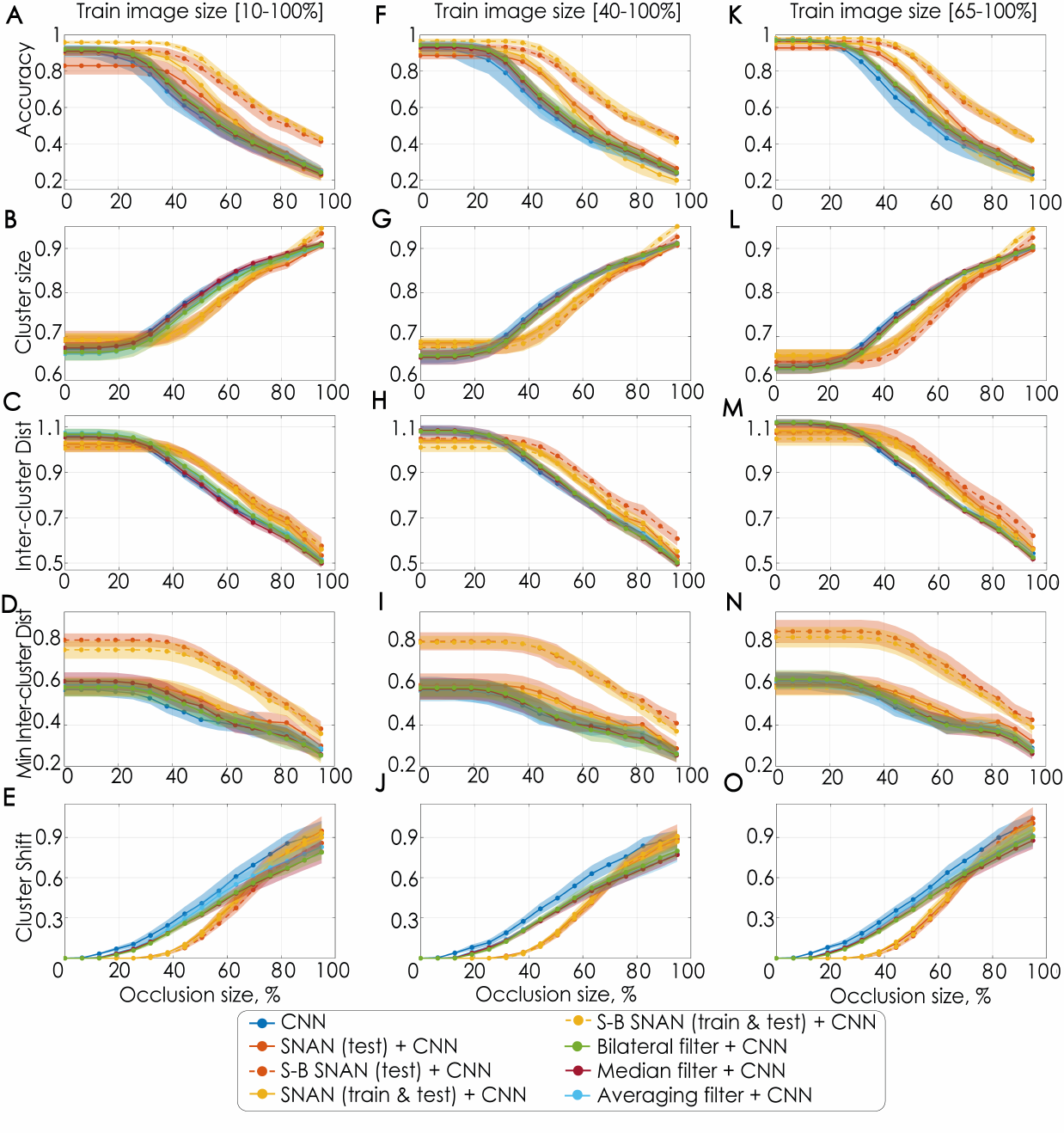}
\caption{Evolution of classification accuracy and feature space properties with increasing occlusion size (percentage of target image area) on the MNIST dataset for CNN-based models trained using different scale augmentation ranges. (A-E) Trained with scale augmentation $X,Y \sim \mathcal{U}$[10\%,100\%], (F-J) Trained with scale augmentation $X,Y \sim \mathcal{U}$[40\%,100\%], (K-O) Trained with scale augmentation $X,Y \sim \mathcal{U}$[65\%,100\%]. (A,F,K) Classification accuracy, (B,G,L) Normalized cluster size of learned representations, $\bar{C_s}$. (C, H, M) Pairwise inter-cluster distance, $\bar{C_d}$. (D, I, N) Nearest-class inter-cluster distance. (E, J, O) Cluster shift. All results are averaged over 10 runs; the solid lines represent the mean and the shaded regions correspond to $\pm$ SD.}
\label{FigS3}
\end{figure}

\begin{figure}
\centering
\includegraphics[width=1\textwidth]{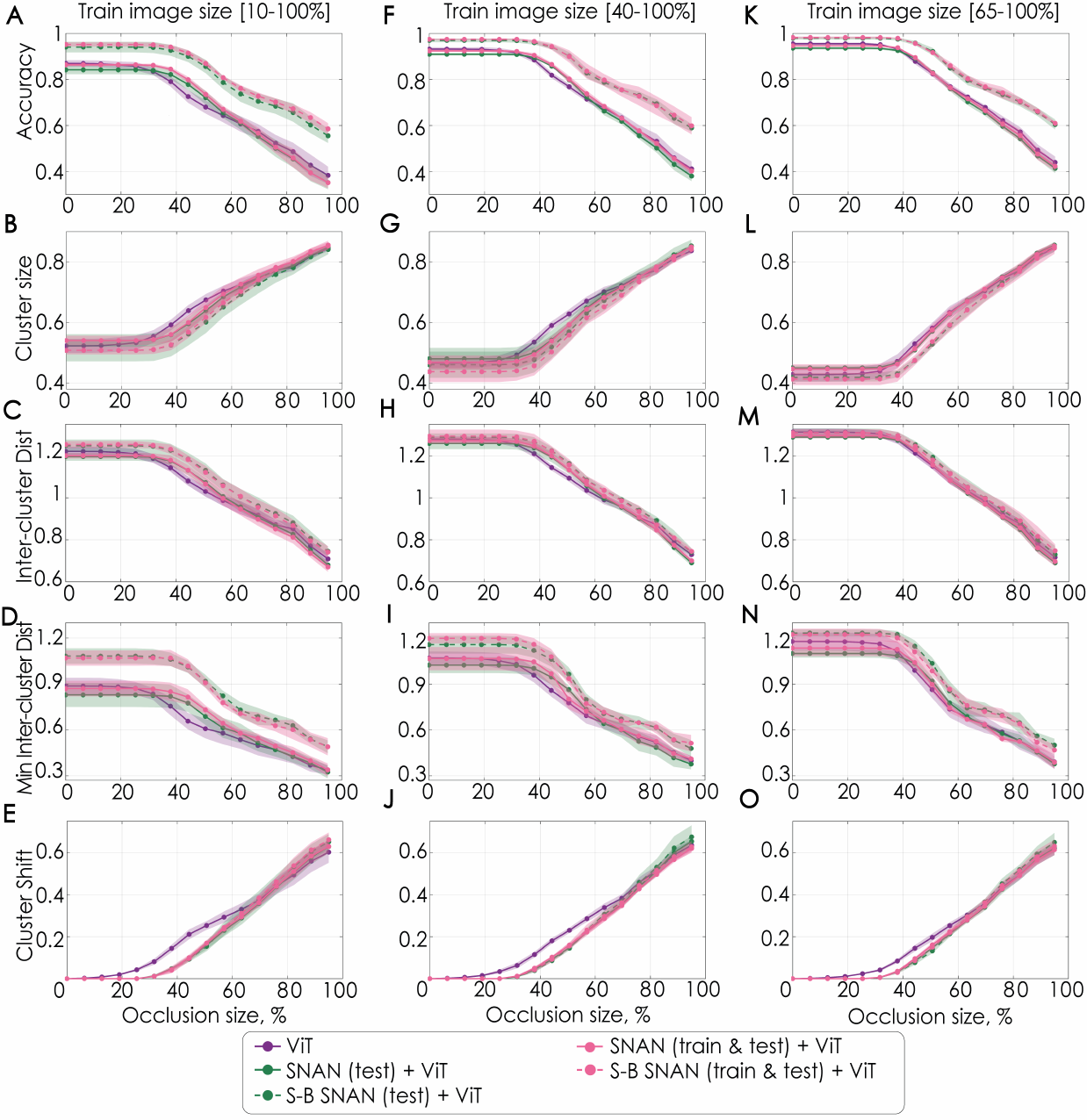}
\caption{Evolution of classification accuracy and feature space properties with increasing occlusion size (percentage of target image area) on the MNIST dataset for ViT-based models trained using different scale augmentation ranges. (A-E) Trained with scale augmentation $X,Y \sim \mathcal{U}$[10\%,100\%], (F-J) Trained with scale augmentation $X,Y \sim \mathcal{U}$[40\%,100\%], (K-O) Trained with scale augmentation $X,Y \sim \mathcal{U}$[65\%,100\%]. (A,F,K) Classification accuracy, (B,G,L) Normalized cluster size of learned representations, $\bar{C_s}$. (C, H, M) Pairwise inter-cluster distance, $\bar{C_d}$. (D, I, N) Nearest-class inter-cluster distance. (E, J, O) Cluster shift. All results are averaged over 10 runs; the solid lines represent the mean and the shaded regions correspond to $\pm$ SD.}
\label{FigS4}
\end{figure}

\begin{figure}
\centering
\includegraphics[width=1\textwidth]{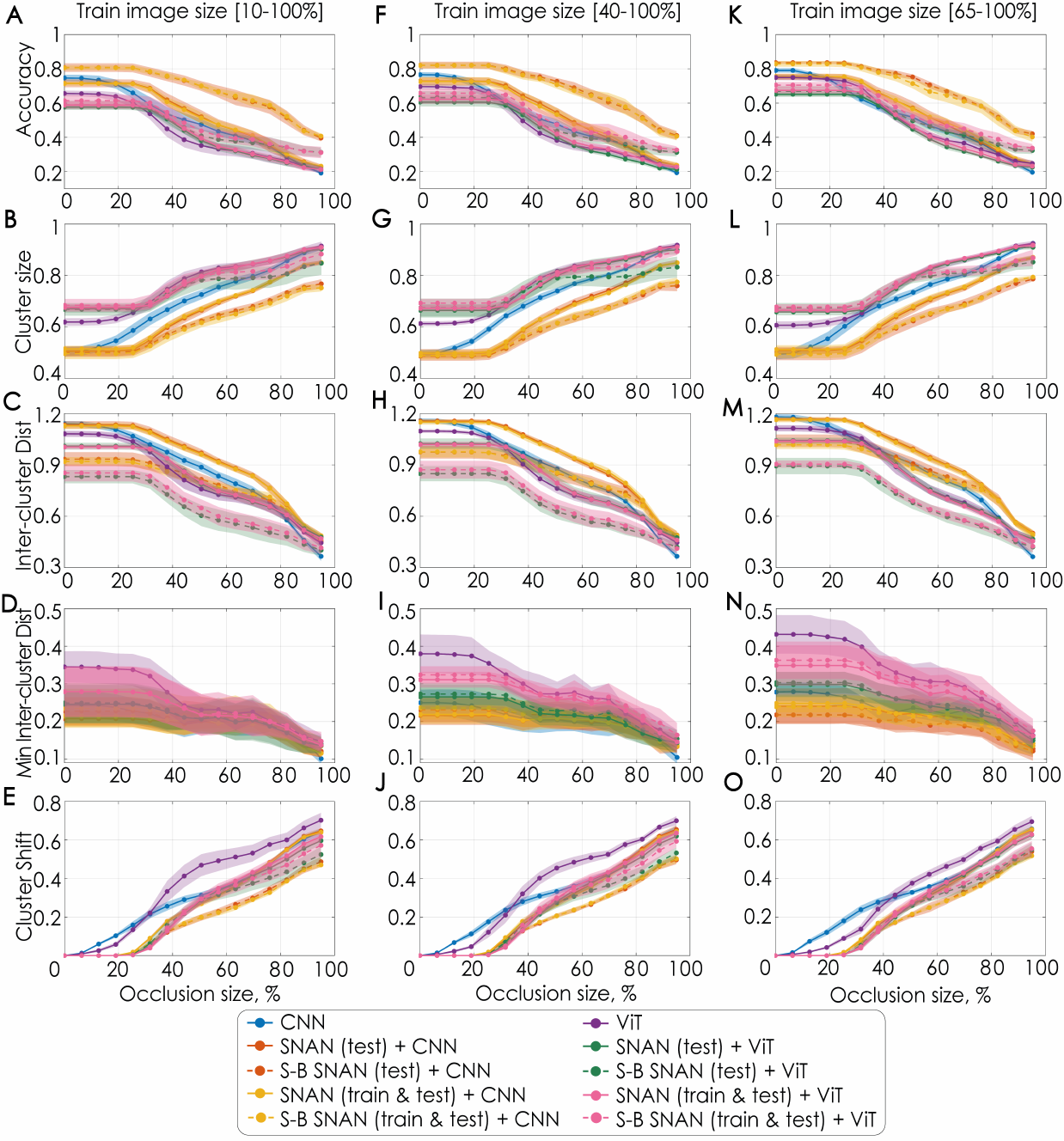}
\caption{Evolution of classification accuracy and feature space properties with increasing occlusion size (percentage of target image area) on the Fashion-MNIST dataset for models trained using different scale augmentation ranges. (A-E) Trained with scale augmentation $X,Y \sim \mathcal{U}$[10\%,100\%], (F-J) Trained with scale augmentation $X,Y \sim \mathcal{U}$[40\%,100\%], (K-O) Trained with scale augmentation $X,Y \sim \mathcal{U}$[65\%,100\%]. (A,F,K) Classification accuracy, (B,G,L) Normalized cluster size of learned representations, $\bar{C_s}$. (C, H, M) Pairwise inter-cluster distance, $\bar{C_d}$. (D, I, N) Nearest-class inter-cluster distance. (E, J, O) Cluster shift. All results are averaged over 10 runs; the solid lines represent the mean and the shaded regions correspond to $\pm$ SD.}
\label{FigS6}
\end{figure}

\end{appendices}

\bibliography{sn-bibliography} 

\end{document}